\documentclass[preprint2,twocolumn,times,tighten]{aastex63}

\usepackage{graphicx,times}
\usepackage{subfigure}
\newcommand{\be}{\begin{equation}}
\usepackage{threeparttable}
\usepackage{booktabs}
\newcommand{\ee}{\end{equation}}
\newcommand{\bea}{\begin{eqnarray}}
\newcommand{\eea}{\end{eqnarray}}
\usepackage{float}
\usepackage{amsmath}
\usepackage{cases}
\usepackage{longtable}
\usepackage{hyperref}
\usepackage{epstopdf}
\usepackage{amsmath,bm}
\usepackage{amssymb}
\usepackage{natbib}
\usepackage{morefloats}
\usepackage{multirow}
\usepackage{array}
\usepackage{verbatim}
\usepackage{color}
\usepackage{CJK}
\usepackage{lineno}
\usepackage{verbatim} 

\begin{document}
\begin{CJK*}{UTF8}{gbsn}

\title{Cosmic ray diffusion in magnetic fields amplified by nonlinear turbulent dynamo}

\email{chao.zhang@ufl.edu;
xusiyao@ufl.edu}

\author[0009-0001-4012-2892]{Chao Zhang}
\affiliation{Department of Physics, University of Florida, 2001 Museum Rd., Gainesville, FL 32611, USA}

\author[0000-0002-0458-7828]{Siyao Xu\footnote{NASA Hubble Fellow}}
\affiliation{Department of Physics, University of Florida, 2001 Museum Rd., Gainesville, FL 32611, USA}

\begin{abstract}
The diffusion of cosmic rays (CRs) in turbulent magnetic fields is fundamental to understand various astrophysical processes. We explore the CR diffusion in the magnetic fluctuations amplified by the nonlinear turbulent dynamo, in the absence of a strong mean magnetic field. Using test particle simulations, we identify three distinct CR diffusion regimes: mirroring, wandering, and magnetic moment scattering (MMS).
With highly inhomogeneous distribution of the dynamo-amplified magnetic fields, we find that the diffusion of CRs is also spatially inhomogeneous.
Our results reveal that lower-energy CRs preferentially undergo the mirror and wandering diffusion in the strong-field regions, and the MMS diffusion in the weak-field regions. The former two diffusion mechanisms play a more important role toward lower CR energies, resulting in a relatively weak energy dependence of the overall CR mean free path.
In contrast, higher-energy CRs predominantly undergo the MMS diffusion, for which the incomplete particle gyration, i.e., the limit case of mirroring, in strong fields has a more significant effect than the scattering by small-scale field tangling/reversal. Compared with lower-energy CRs, they are more poorly confined in space, and their mean free paths have a stronger energy dependence. 
We stress the fundamental role of magnetic field inhomogeneity of nonlinear turbulent dynamo in 
causing the different diffusion behavior of CRs compared to that in sub-Alfv\'enic MHD turbulence.

\end{abstract}

\section{Introduction}
Diffusion of cosmic rays (CRs) in turbulent magnetic fields is the key to understanding many important processes in astrophysics and space physics, such as nonthermal emission from galaxies and galaxy clusters (\cite{brunetti2011acceleration}; \cite{feretti2012clusters}; \cite{krumholz2020cosmic}; \cite{pfrommer2022simulating}; \cite{yang2023faraday}), and high-energy neutrino production from active galactic nuclei (\cite{becker2022propagation}). 
Its theoretical study is of particular importance for the multi-messenger astronomy 
(\cite{halzen2003multi}; \cite{de2023cosmic}) 
and understanding recent CR observations that cannot be explained by the existing theoretical paradigm of CR diffusion (e.g., \cite{gabici2019origin}; \cite{evoli2020ams}; \cite{amato2021particle}; \cite{fornieri2021theory}; \cite{hopkins2022standard}). 

Most earlier studies on CR diffusion (e.g., \cite{chandran2000mirrorconfinement}; \cite{yan2002scatteringanisotropy}; \cite{yan2004cosmic}; \cite{beresnyak2011numerical}; \cite{xuyan2013}; \cite{cohet2016cosmic}; \cite{yue2022superdiffusion}; \cite{bustard2023cosmic}) focus on the regime of sub-Alfv\`enic magnetohydrodynamic (MHD) turbulence with a strong mean field component (\cite{LV99reconnection}; \cite{maron2001simulations}; \cite{cho2002compressible}), which can be applied to the interstellar medium of spiral galaxies with large-scale magnetic fields amplified by the mean field dynamo (\cite{vishniac2001magnetic};  \cite{schlickeiser2016cosmic}; \cite{brandenburg2018advances}; \cite{commerccon2019cosmic}). 
For the gyroresonant scattering of CRs by small-amplitude magnetic fluctuations, there is a  long-standing $90^\circ$ problem of the quasi-linear theory (\cite{jokipii1966cosmic}; \cite{schlickeiser1998quasi}).
Particularly, mirroring by magnetic compressions in MHD turbulence presents a natural solution to the $90^\circ$ problem (\cite{cesarsky1973mirror}; \cite{xu2020trapping}). The corresponding mirror diffusion has been recently proposed by \cite{lazarian2021mirroring} (LX21 henceforth) and numerically demonstrated in sub-Alfv\'enic MHD turbulence (\cite{zhang2023numerical} ZX23 henceforth; \cite{barreto2024cosmic}). This new diffusion mechanism can effectively confine CRs in the vicinity of their sources (e.g., \cite{xu2021mirror_supernova}) and account for the suppressed diffusion indicated by recent gamma-ray observations (e.g., \cite{torres2010gev}; \cite{abeysekara2017hwc}).

In the absence of strong mean magnetic field in, e.g, the intracluster medium (ICM), magnetic fluctuations can be amplified by the ``small-scale" turbulent dynamo on scales smaller than the turbulence injection scale (\cite{kazantsev1968enhancement}; \cite{kulsrud1992spectrum};  \cite{brandenburg2005astrophysical}; \cite{xu2016turbulent}; \cite{seta2021saturation}), with folded magnetic structure seen in the kinematic stage at a large magnetic Prandtl number (\cite{schekochihin2004simulations}) and MHD turbulence developed up to the energy equipartition scale in the nonlinear stage (\cite{cho2009growth}; \cite{beresnyak2012universal}; \cite{xu2016turbulent}).
Different diffusion mechanisms of CRs have been proposed and studied in dynamo-amplified magnetic fields. Recent studies by, e.g., \cite{lemoine2023particle} and \cite{kempski2023cosmic}, found that unlike the gyroresonant scattering in sub-Alfv\'enic turbulence, low-energy CRs can undergo significant scattering by weak but highly curved magnetic fields in turbulent dynamo simulations. Moreover, at the final energy saturation of the nonlinear dynamo, the wandering of strong amplified magnetic fields gives rise to the effective diffusion of CRs moving along the field lines, with the effective mean free path determined by the energy equipartition scale (\cite{brunetti2007compressible}).

Since CR diffusion strongly depends on the properties of turbulent magnetic fields, and most earlier studies focus on the regime of sub-Alfv\`enic turbulence, in this study, we will focus on investigating the effect of dynamo-amplified turbulent magnetic fields on CR diffusion.
We will examine the mirror diffusion (LX21) in dynamo-amplified magnetic fields, which has only been numerically tested in sub-Alfv\'enic turbulence (ZX23, \cite{barreto2024cosmic}). In addition, we will also examine the scattering diffusion (\cite{lemoine2023particle}; \cite{kempski2023cosmic}) and the wandering diffusion (\cite{brunetti2007compressible}). The wandering diffusion hasn't been numerically tested before. We will compare their relative importance in affecting the diffusion of CRs at different energies.
The outline of this paper is as follows. In Section \ref{sec:numerical}, we describe the numerical methods for simulating dynamo-amplified turbulent magnetic fields and diffusion of test particles (i.e., CRs). In Section \ref{sec:results}, we study in detail the diffusion regimes of CRs at different energies, and measure their energy-dependent mean free paths. In Section \ref{sec:discussion}, we compare our results with previous studies and discuss implications on diffusion of CR electrons in the ICM. In Section \ref{sec:conclusion}, we present our conclusions.

\begin{table*}[t]
\centering
\caption{Parameters of the dynamo simulation at the nonlinear saturation. }
\centering    
\begin{tabular}{|c c c c c c c c|} 
 \hline
      $M_A$ & $M_s$ & $\beta$ & $ B_\text{rms}/ B_0$ &  Resolution & $L_\text{inj}$ & $l_\text{diss}$ & $l_\text{eq}$ \\
 \hline\hline
   1.19  & 0.84   & 4.03 &  $\mathcal{O}(10^1)$ & $792^3$  & $400$ & $\approx 16$ &  $\approx 160$    \\
 \hline
\end{tabular}
 \label{table:dataM1}
\end{table*}

\begin{figure}[t]
  \centering
    \hspace*{-.5cm}\includegraphics[width=0.5\textwidth]{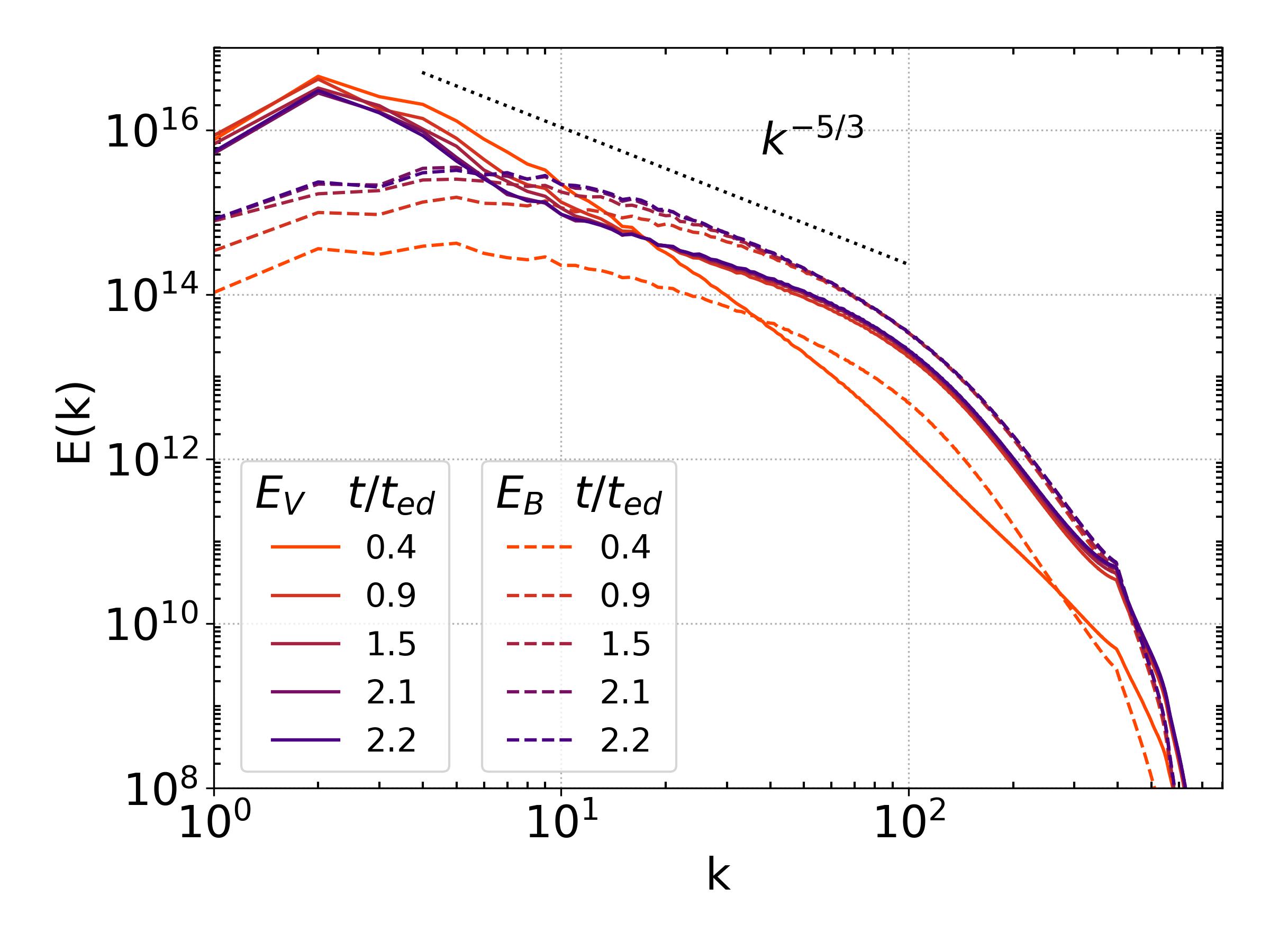}
    \hspace*{0.cm}
  \caption{ Turbulent kinetic ($E_V$) and magnetic ($E_B$) energy spectrum measured at different times. The snapshot used for test particle simulation corresponds to $t/t_\text{ed}=2.2$ when the nonlinear saturation of the dynamo is reached. }
  \label{fig:spectrum}
\end{figure}

\section{Numerical method}
\label{sec:numerical}

\begin{figure*}[htb!]
\gridline{\fig{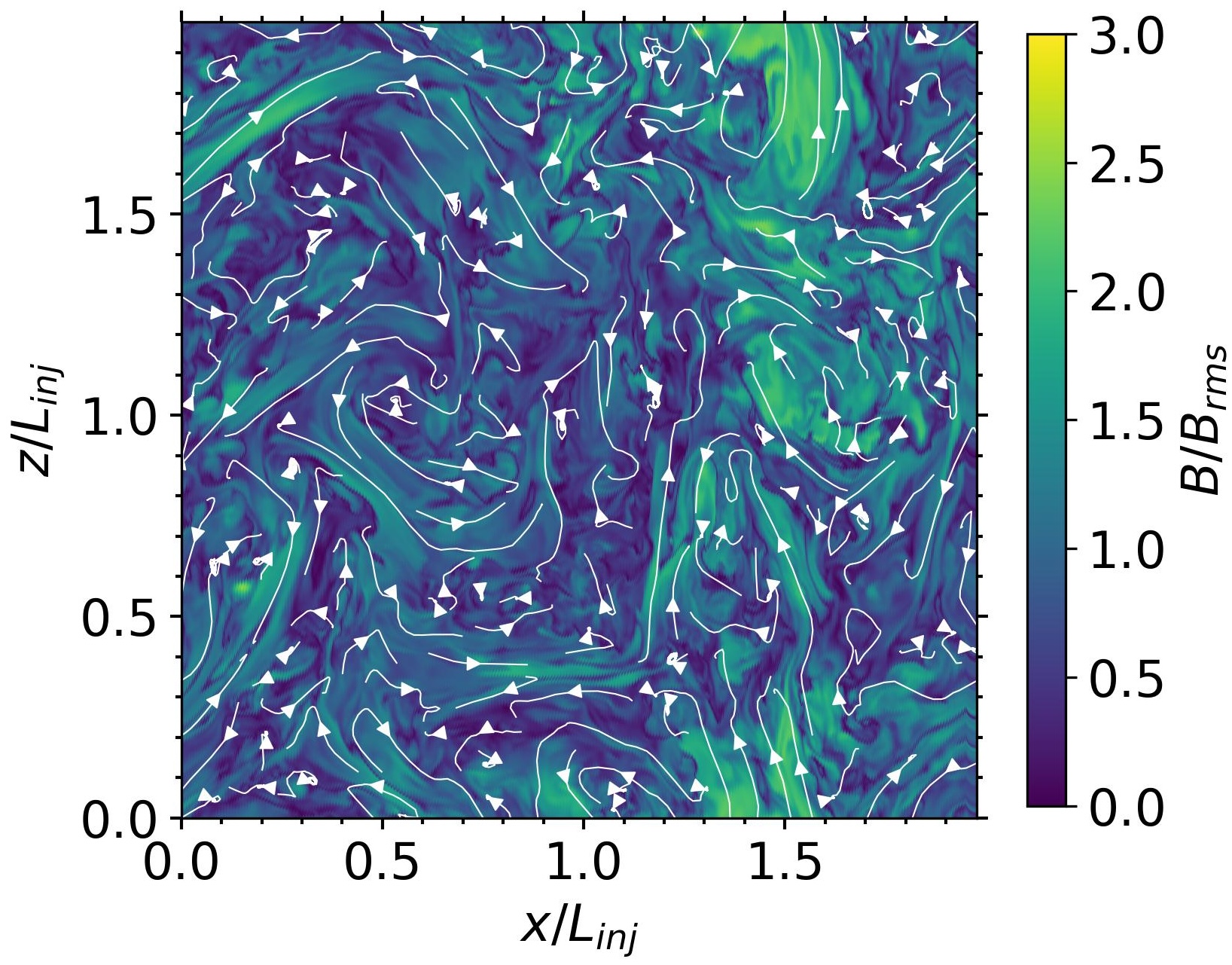}{0.45\textwidth}{(a)}
          \fig{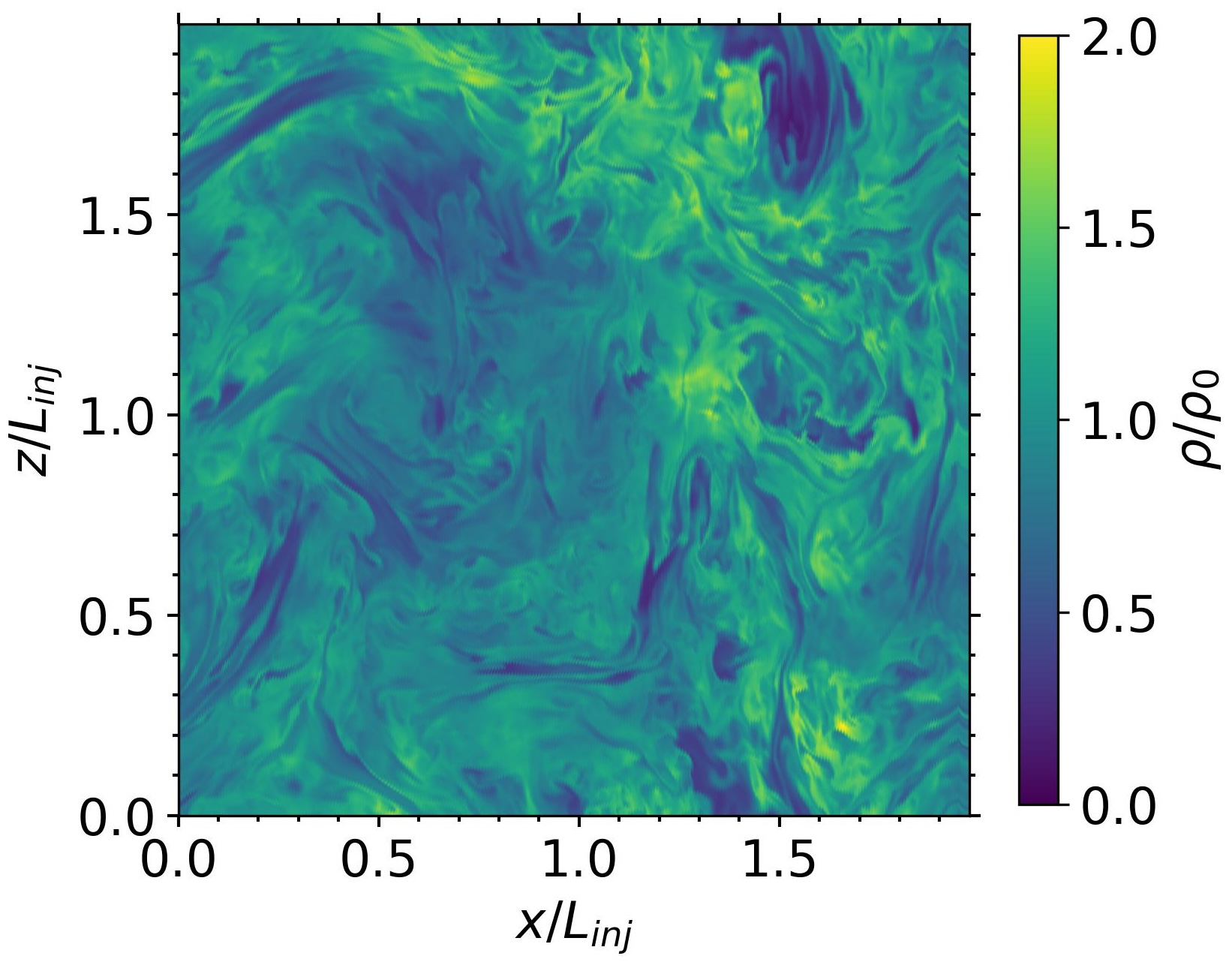}{0.45\textwidth}{(b)}
          }
\caption{ Two-dimensional slices of (a) magnetic fields and (b) gas density distributions at $y/L_\text{inj}=1.0$ taken from the 3D dynamo simulation at the final nonlinear saturation. Streamlines in (a) represent the direction of local magnetic fields in the slice plane. $\rho_0$ is the mean gas density.}
\label{fig:BsliceM1}
\end{figure*}

\begin{figure}[htb!]
  \centering
    \hspace*{-0cm}\includegraphics[width=0.45\textwidth]{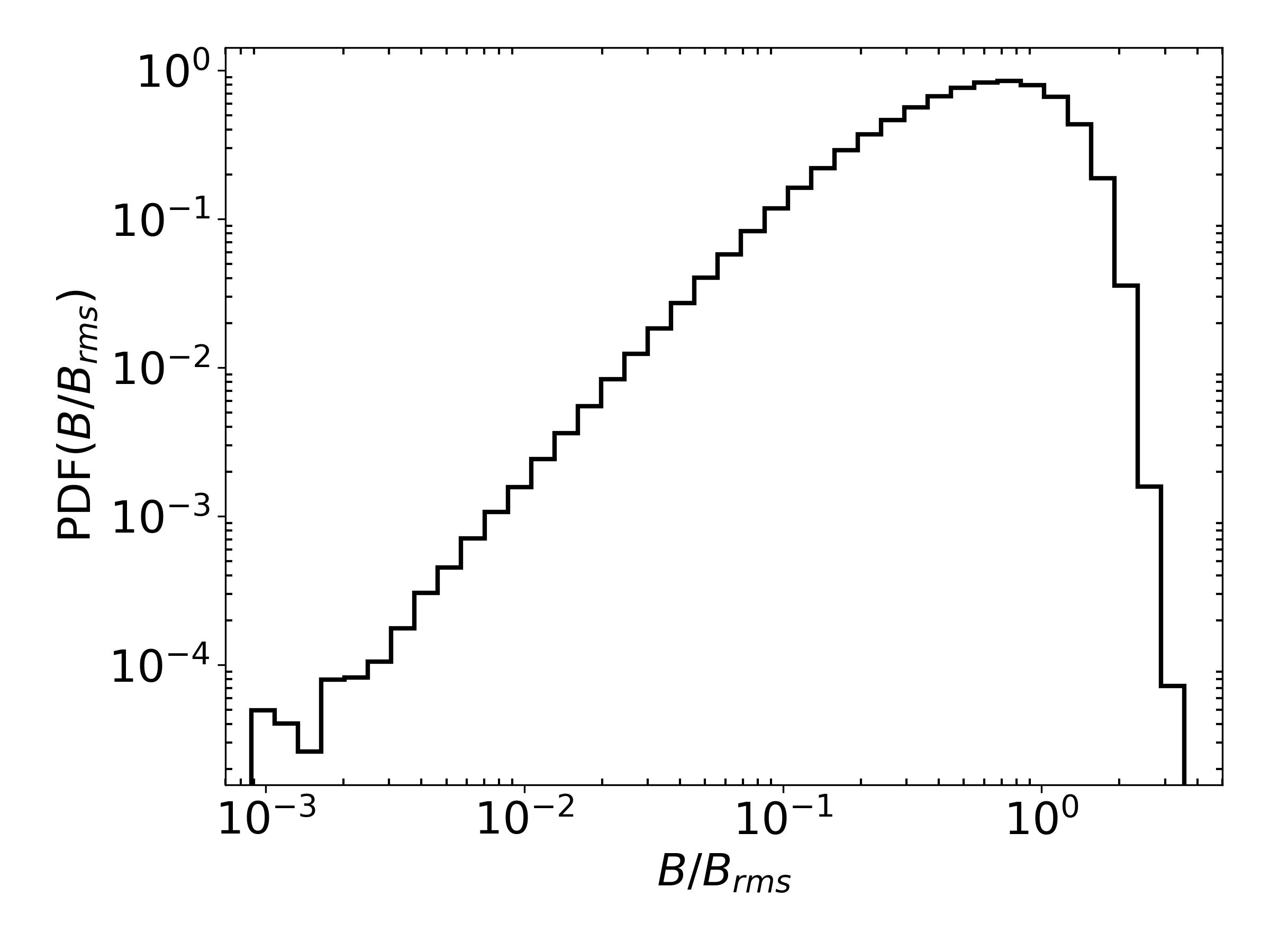}
  \caption{ Probability density function (PDF) of magnetic field strength $B/B_\text{rms}$ of the dynamo simulation at the nonlinear saturation.}
  \label{fig:BdisM1}
\end{figure}

This work focuses on the diffusion of CRs in nonrelativistic turbulence, with magnetic fluctuations amplified by turbulent dynamo. First we carry out the dynamo simulation with an initially weak magnetic field. Then we take a snapshot of the dynamo-amplified magnetic fields at later time of the nonlinear stage of dynamo when the magnetic energy is fully saturated. Since CRs have speed (basically the speed of light) much higher than the turbulent speed, we can treat turbulent magnetic fields as a static background. Then we inject test particles that represent CRs in the turbulent magnetic fields and numerically integrate their equation of motion to obtain the trajectories, similar to the procedure in, e.g., \cite{xuyan2013} and ZX23. 

\subsection{Turbulent Dynamo Simulation}
\label{sec:MHD}
The simulation of turbulent dynamo is performed with \textsc{Athena++} that solves isothermal ideal compressible MHD equations in a periodic cubic box (\cite{stone2020athena++}). The driving of turbulence is continuous over time in Fourier space over a range of scales corresponding to wavenumber $k=2$ to 4 with a power law $k^{-5/3}$. The maximal driving energy is thus at $k=2$ (half-box size), corresponding to the injection scale $L_\text{inj}$. The driving is fully solenoidal, following an Ornstein-Uhlenbeck process with correlation time chosen to be the largest-eddy turnover time $t_\text{ed}=L/2V_L$, where $L$ is the box size and $V_L$ is the turbulent injection velocity approximately equal to the root-mean-square (RMS) magnitude of the velocity field. 
The sonic Mach number is defined as $M_s=V_L/c_s$, where $c_s$ is the isothermal sound speed. 
The Alfv\'en Mach number $M_A$ is measured as the ratio of the injection velocity $V_L$ to the Alfv\'en velocity $V_A$, i.e., $M_A=V_L/V_A$. The Alfv\'en velocity is determined by $V_A=B_\text{rms}/\sqrt{4\pi \rho_0}$ where $B_\text{rms}$ is the RMS magnetic field strength and $\rho_0$ is the mean gas density.
A uniform mean-field $\mathbf{B}_0=[0,0,B_0]$ is initialized in the z-direction of the box. $B_0$ is very weak for dynamo simulation, and accordingly $\beta$ is initially much greater than unity, where the plasma beta $\beta=2(c_s/V_A)^2$ is the gas pressure over magnetic pressure.
$B_\text{rms}$ significantly increases during the dynamo process, and eventually reaches a value significantly larger than $B_0$ at the final saturation.
 
The parameters of the extracted snapshot of the dynamo simulation at the nonlinear saturation are summarized in Table \ref{table:dataM1}. The resolution is $792^3$ in grid unit and the injection scale $L_\text{inj}$ is approximately at $400$ grid. The numerical dissipation scale $l_\text{diss}$ appears at $k\approx 50$, corresponding to $\approx 16$ grids.
The magnetic and turbulent kinetic energy spectra at different evolution times are shown in Fig. \ref{fig:spectrum}. At the scales below the energy equiparition scale $l_\text{eq}\approx 160$ grid (corresponding to the peak of magnetic energy spectrum at $k\approx 5$), the magnetic and turbulent energies are in equipartition.
Note that the gap between magnetic and kinetic energy at larger scales $l>l_\text{eq}$ is commonly seen in small-scale dynamo simulations that do not include magnetic helicity (\cite{shapovalov2011simulations}).
In Fig. \ref{fig:BsliceM1}, we show two 2D slices of magnetic field strength and gas density distributions at $y=1.0L_\text{inj}$ taken from the 3D simulation. The streamlines in Fig. \ref{fig:BsliceM1} (a) illustrate the magnetic field lines. We see that the magnetic field distribution is highly inhomogeneous, with both strong-field regions with smooth field lines and weak-field regions with tangled field lines. In the strong-field regions, magnetic fields are amplified by turbulent stretching, and the magnetized turbulence is expected to have similar properties as trans-Alfv\'enic MHD turbulence (\cite{xu2016turbulent}). In the weak-field regions, magnetic fields are passively advected by turbulent motions, and the turbulence has hydro-like properties. 
In Appendix \ref{sec:bslice}, the distributions of magnetic fields with $B<B_\text{rms}$ and $B\geq B_\text{rms}$ are separately presented in Fig. \ref{fig:Bslice_cut}. Comparing the magnetic field strength and gas density distributions in Fig. \ref{fig:BsliceM1}, the anticorrelation between the magnetic field and gas density is seen in some regions. In addition, Fig. \ref{fig:BdisM1} shows the probability density function (PDF) of magnetic field strength $B/B_\text{rms}$, as a relatively broad distribution ranging from $B/B_\text{rms}\approx 10^{-3}$ to $3$. 

We see that unlike the sub-Alfv\'enic MHD turbulence, dynamo-amplified magnetic fields are highly inhomogeneous, with a broad distribution of magnetic field strengths. We will study its effect on the CR diffusion in Section \ref{sec:results}.

\subsection{Test Particle Simulations}
\label{sec:particle}
Test particles are injected in the dynamo-amplified magnetic fields with their trajectories obtained by integrating the following equations of motion,
\begin{equation}
    \frac{d\mathbf{u}}{dt} = \frac{q}{\gamma mc}\mathbf{u}\times \mathbf{B(\mathbf{r})}~,\label{eq:lorentz}
\end{equation}
and
\begin{equation}
    \frac{d\mathbf{r}}{dt} = \mathbf{u}~,\label{eq:velocity}
\end{equation}
for particles with Lorentz factor $\gamma$, mass $m$, charge $q$ and velocity $\mathbf{u}$. $c$ is the speed of light and $t$ is the time. The local magnetic field $\mathbf{B}(\mathbf{r})$ at particle position $\mathbf{r}$ along the particle trajectory is interpolated using a cubic spline constructed from the magnetic fields at the $10^3$ neighbouring points.
The Bulirsch-Stoer method (\cite{press1988numerical}) is adopted for solving Eq. \ref{eq:lorentz} and Eq. \ref{eq:velocity} with an adaptive time step, which provides a high accuracy for the particle trajectory integration with a relatively low computational effort. 
The energy of CRs is quantified by the ratio $r_L(B_\text{rms})/L_\text{inj}$ with the Larmor radius $r_L(B_\text{rms})$ defined using $B_\text{rms}$ as
\begin{equation}
    r_L(B_\text{rms}) = \frac{\gamma mc^2}{q B_\text{rms}}~.
    \label{eq:rL}
\end{equation}
Particle can have a different local Larmor radius $r_L$ at a position $\mathbf{r}$ depending on the local magnetic field strength $B(\mathbf{r})$, which is evaluated by $r_L(B)=r_L(B_\text{rms})B_\text{rms}/B(\mathbf{r})$. Due to the broad distribution of $B$ for dynamo-amplified magnetic fields (see Fig. \ref{fig:BdisM1}), the distribution of $r_L(B)$ of CRs is also broad. Throughout the test particle simulations, we set the range of CR energies for the value of local $r_L(B)$ to be greater than the grid size and the gyromotion to be always resolved. 
The injection scale $L_\text{inj}$ is used to normalize Larmor radius as $L_\text{inj}$ is well determined in our simulations and is convenient for astrophysical applications.
The gyrofrequency is defined as $\Omega = u/r_L(B_\text{rms})$. 
At every time step of particle motion, we measure the pitch-angle cosine $\mu = \cos \theta~$,
where $\theta$ is the pitch-angle, which is the angle between $\mathbf{B}(\mathbf{r})$ and $\mathbf{u}$. The magnetic moment (i.e., the first adiabatic invariant) of CRs is defined as
\begin{equation}
    M= \frac{\gamma m u_\perp^2}{2B} ~,\label{eq:adiabatic}
\end{equation}
where $u_\perp=u(1-\mu^2)^{1/2}$ is the particle velocity perpendicular to the local magnetic field. Each test particle simulation contains around 2000 particles of the same energy, which provides a sufficiently large sample size for the results to be convergent (\cite{xuyan2013}; ZX23). All test particles are injected at random initial positions with random $\mu$. 

\section{CR diffusion}
\label{sec:results}
In this section, we investigate three distinct CR diffusion regimes that we identify and define as mirroring, wandering, and magnetic moment scattering (hereafter MMS). Their characteristics and identification methods are discussed in Section \ref{sec:CRdiffusion}. Their time fractions at different CRs energies are presented in Section \ref{sec:statistical}. In Section \ref{sec:mfp}, we measure the overall mean free paths (MFPs) of CRs at different energies.

\subsection{CR diffusion regimes}
\label{sec:CRdiffusion}

\begin{figure}[t]
  \centering
    \hspace{-0.5cm}\includegraphics[width=0.48\textwidth]{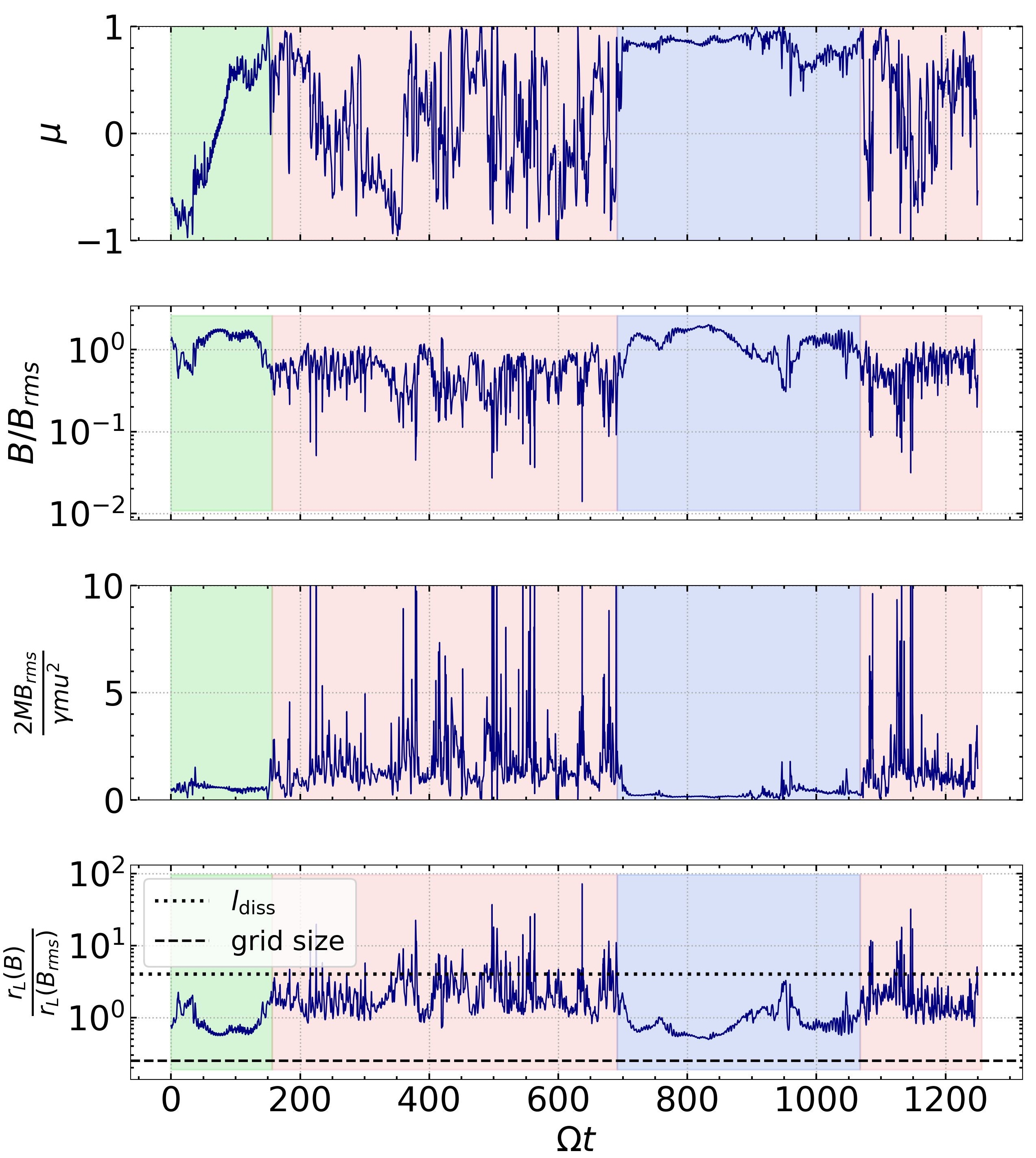}
  \caption{ Time evolution of $\mu$, $B/B_\text{rms}$, $2MB_\text{rms}/\gamma mu^2$ and $r_L(B)/r_L(B_\text{rms})$ for a test particle with $r_L(B_\text{rms})=0.01L_\text{inj}$. The regimes of mirroring, wandering and MMS diffusion are shaded by green, blue, and red, respectively. The numerical dissipation scale $l_\text{diss}$ and grid size are indicated by horizontal dotted and dashed lines in the last panel.}
  \label{fig:CRM1_4}
\end{figure}

\begin{figure}[t]
  \centering
    \hspace{-0.5cm}\includegraphics[width=0.48\textwidth]{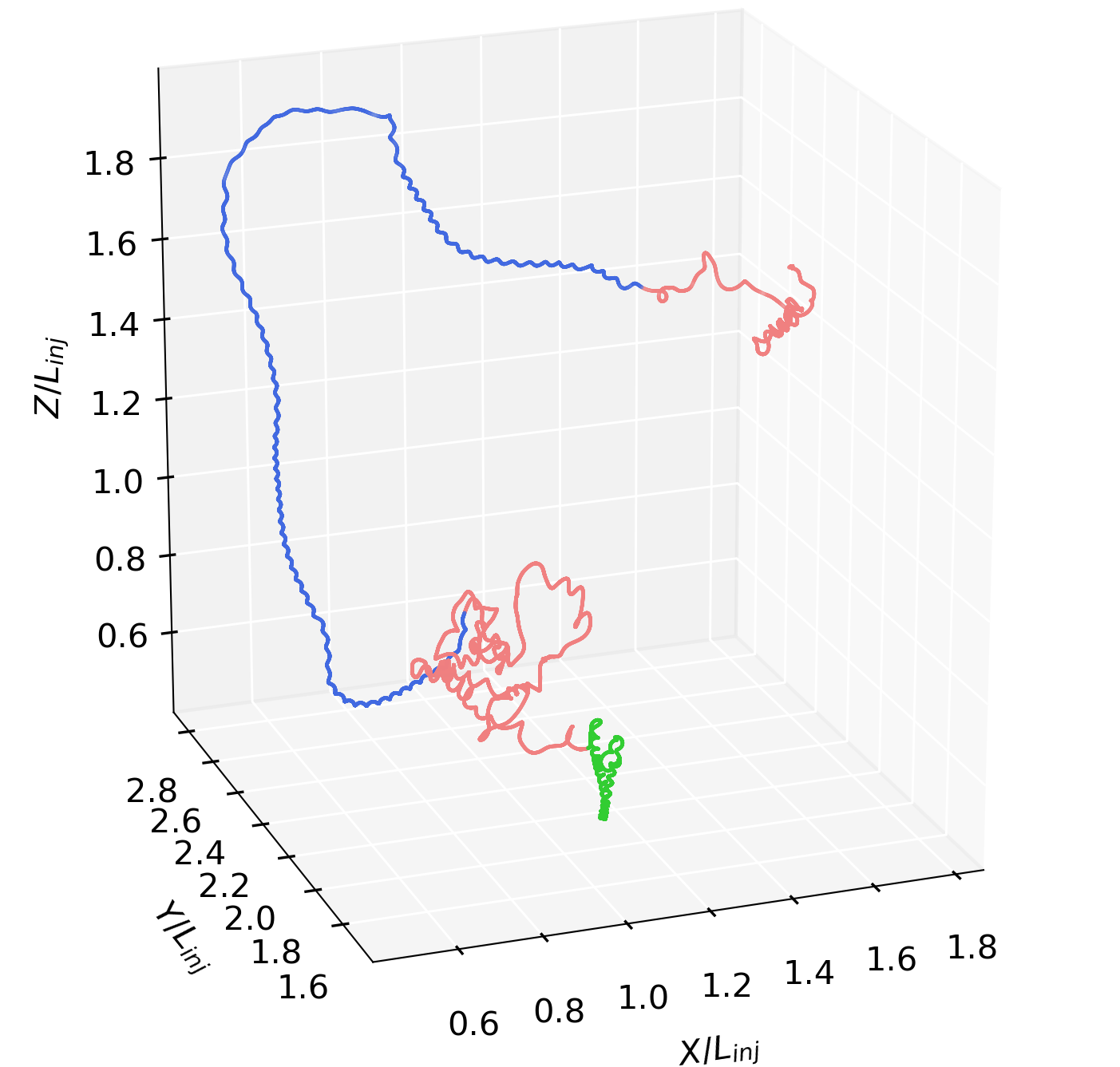}
  \caption{Trajectory for the test particle shown in Fig. \ref{fig:CRM1_4}. Segments of the trajectory where the particle undergoes mirroring, wandering and MMS are color-coded by green, blue and red. }
  \label{fig:CRM1_4_3d}
\end{figure}

\begin{figure*}[htb!]
\gridline{
          \fig{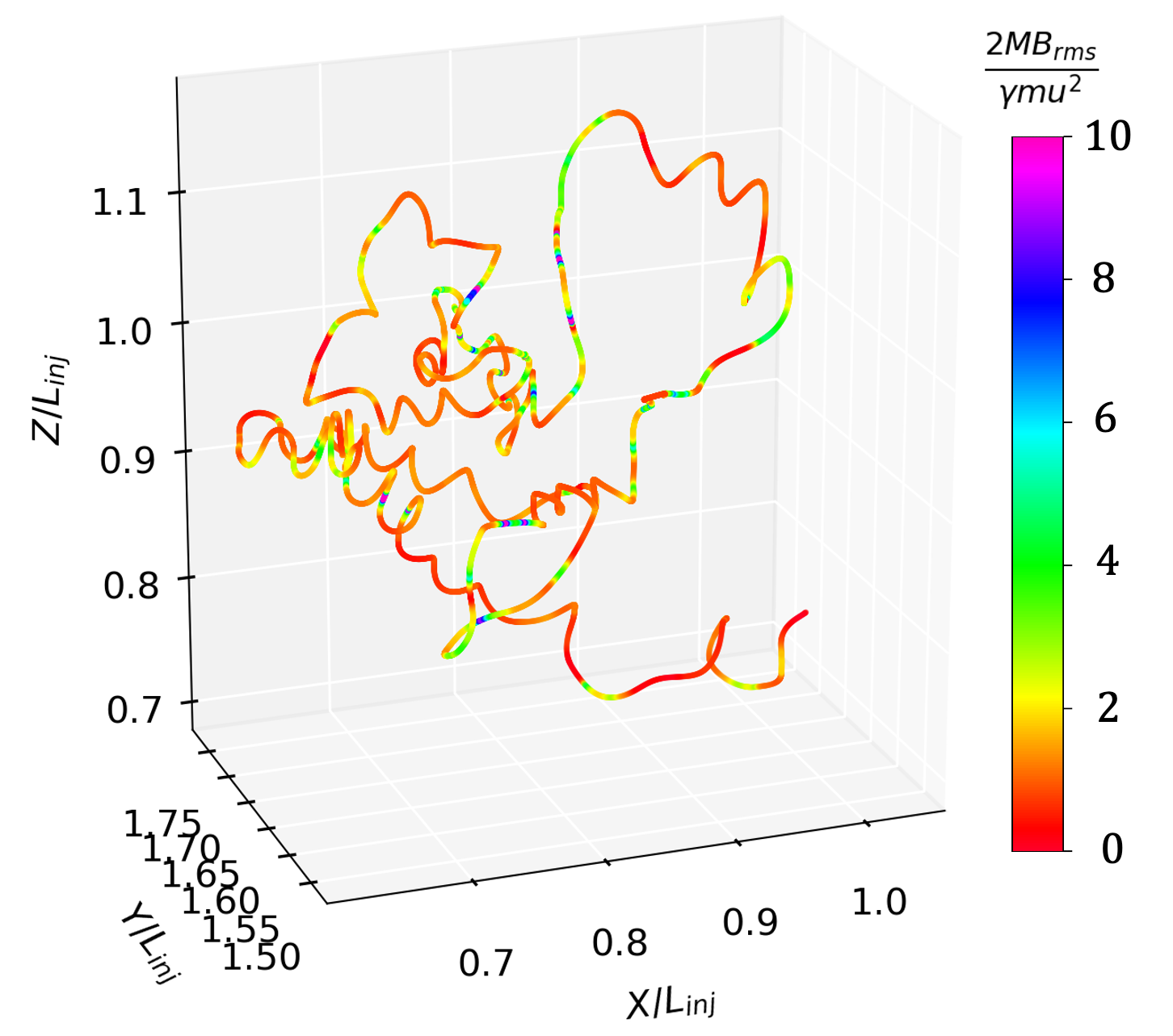}{0.45\textwidth}{(a)}
          \fig{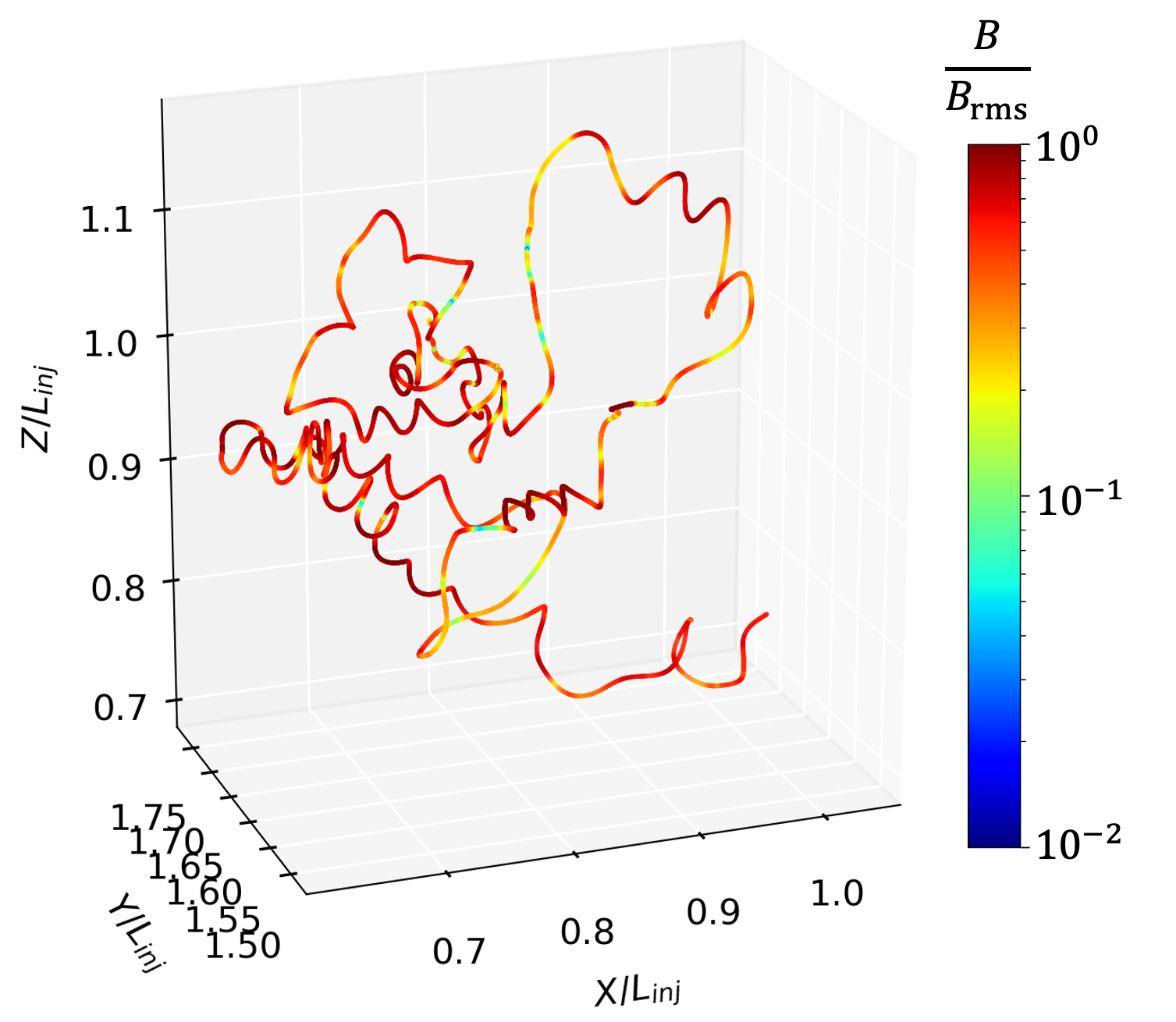}{0.45\textwidth}{(b)}
          }
\caption{ (a) Zoom-in on one of the MMS regimes from the trajectory in Fig. \ref{fig:CRM1_4_3d}, color-coded with the magnetic moment. (b) Same as (a) but color-coded with the magnetic field strength. 
The sharp turns in the trajectory
are mostly caused by incomplete gyrations with constant $M$, i.e., the limit case of mirroring, in stronger fields.}
\label{fig:CRM1_4_MMS}
\end{figure*}

(i) Mirror diffusion. Turbulent compression of magnetic fields leads to the formation of magnetic mirrors that can reflect CRs with sufficiently large pitch angles (\cite{xu2020trapping}). Due to perpendicular superdiffusion of turbulent magnetic fields (\cite{xuyan2013}; \cite{lazarian2014superdiffusion}; \cite{yue2022superdiffusion}), these mirroring CRs are not trapped but move diffusively along the turbulent magnetic fields, which is the mirror diffusion (LX21; ZX23). Mirroring CRs have $r_L(B)$ smaller than the mirror size and their pitch angles satisfy $\mu<\mu_\text{max}$, where $\mu_\text{max}$ corresponds to the smallest pitch angle for mirroring, which depends on the amplitude of magnetic compressions and the efficiency of pitch angle scattering at a given CR energy (LX21). Mirroring CRs have no stochastic change of $\mu$, and their $M$ is conserved. In ZX23, it is numerically demonstrated that the mirror diffusion is much slower than the diffusion associated with gyroresonant scattering and can effectively confine CRs in space.

(ii) Wandering diffusion. For CRs that do not undergo either mirroring or efficient scattering, they can follow the turbulent magnetic field lines with $\mu$ remaining constant.
The dynamo-amplified strong magnetic fields change their orientations in a random walk manner over $l_\text{eq}$. Therefore, CRs moving along the field lines have an effective mean free path determined by $l_\text{eq}$, which was proposed by \cite{brunetti2007compressible} as a diffusion mechanism. We term it ``wandering diffusion" as it originates from the field line wandering in turbulent dynamo. The wandering mechanism has already been considered as a mechanism for spatially confining particles in the ICM (\cite{brunetti2007compressible}) and in galaxies (e.g., \cite{krumholz2020cosmic}; \cite{xu2022turbulent}).

(iii) MMS diffusion. CRs with small pitch angles in the MHD turbulence with strong mean magnetic field component are subject to gyroresonant scattering (\cite{schlickeiser2013cosmic}). 
However, the quasi-linear approximation for describing the gyroresonant scattering is no longer valid in MHD turbulence with magnetic fluctuation comparable to the mean magnetic field (\cite{yan2008cosmic}; \cite{xu2018TTD_anisotropy}) and dynamo-amplified magnetic fields with $\delta B \gg B_0$ (\cite{lemoine2023particle}; \cite{kempski2023cosmic}). 
The scattering by large magnetic fluctuations is expected to be characterized by significant variations of magnetic moment, i.e., $\Delta M/M> 1$, hence it is termed magnetic moment scattering (MMS, \cite{delcourt2000magnetic}). MMS has been studied for charged particles in a field reversal in, e.g., the earth's geomagnetic tail (\cite{chen1992nonlinear}). With the weak tangled magnetic fields present in the dynamo process, similar scattering with the local $r_L(B)$ comparable to the curvature radius of the magnetic field is also seen in dynamo simulations (\cite{lemoine2023particle}; \cite{kempski2023cosmic}).

We first analyze the diffusion behavior of  low-energy CRs with $r_L(B_\text{rms}) < l_\text{diss}$. As an example, Fig. \ref{fig:CRM1_4} shows the time evolution of $\mu$, $B(\mathbf{r})/B_\text{rms}$, normalized magnetic moment $2MB_\text{rms}/\gamma mu^2$ and normalized local Larmor radius $r_L(B)/r_L(B_\text{rms})$ over normalized time $\Omega t$, where $\Omega$ is the gyrofrequency defined with $B_\text{rms}$, for a test particle with $r_L(B_\text{rms})=0.01L_\text{inj}$. From these measured quantities, we can identify three regimes of diffusion, including mirroring, wandering and MMS diffusion, shaded by green, blue and red respectively. The mirroring regime is characterized by crossings at $90^\circ$, i.e., $\mu=0$, and constant $M$. Wandering is characterized by constant large $\mu$ and constant $M$. We find that mirroring and wandering preferentially take place when the local magnetic field is so strong that the corresponding $r_L(B)$ becomes smaller than $l_\text{diss}$ and scattering is not expected. By contrast, MMS shows a distinctive feature that all four quantities undergo large stochastic variations. It preferentially takes place when the magnetic field is relatively weak with local drops of magnetic field strengths. $r_L(B)$ can exceed $l_\text{diss}$ when the local magnetic field is sufficiently weak. Note that this example is not for representing the time fraction of each diffusion regime (see Section \ref{sec:statistical}), but for illustrating their characteristics that are used for their identification. In addition, we present the corresponding particle trajectory in Fig. \ref{fig:CRM1_4_3d}, where the color-coding for the three diffusion regimes is the same as in Fig. \ref{fig:CRM1_4}. Similar to our finding in ZX23, the mirroring causes the particle to reverse its moving direction and thus effective confinement in space. When moving along the strong and coherent field line, the diffusion is purely governed by the field line wandering. 
Fig. \ref{fig:CRM1_4_MMS} (a) and (b) show the zoom-ins of the particle trajectory in the MMS regime, color-coded by the magnetic moment and magnetic field strength, respectively. By careful inspection, we find that due to the inhomogeneous distribution of magnetic fields (see also Fig. \ref{fig:Bslice_cut} (a)), the particle motion alternates between gyrations (including complete and incomplete ones) around stronger fields with constant $M$ and random behavior in weak fields with changing $M$. We note that unlike the mirroring, the crossing at $\mu=0$ caused by scattering does not correspond to reversing but a random and milder change of the particle's moving direction, as the local direction of the tangled field is poorly defined. As a result, we see a relatively smooth trajectory in weak fields. By contrast, the sharp turns in the trajectory are mostly caused by incomplete gyrations in stronger fields.
With constant $M$ in the presence of stronger magnetic fields, the incomplete gyration leading to an abrupt change of a particle's moving direction resembles a mirroring event, but we consider it as a limit case of mirroring as the mirror size is comparable to the local $r_L(B)$.
Compared to the mirroring and MMS, the CR in the wandering regime travels over a larger distance and has a faster diffusion. The relatively poor confinement by wandering is due to the large correlation length ($\sim l_\text{eq}$) of dynamo-amplified magnetic fields.

\begin{figure}[t!]
  \centering
    \hspace{-0.2cm}\includegraphics[width=0.48\textwidth]{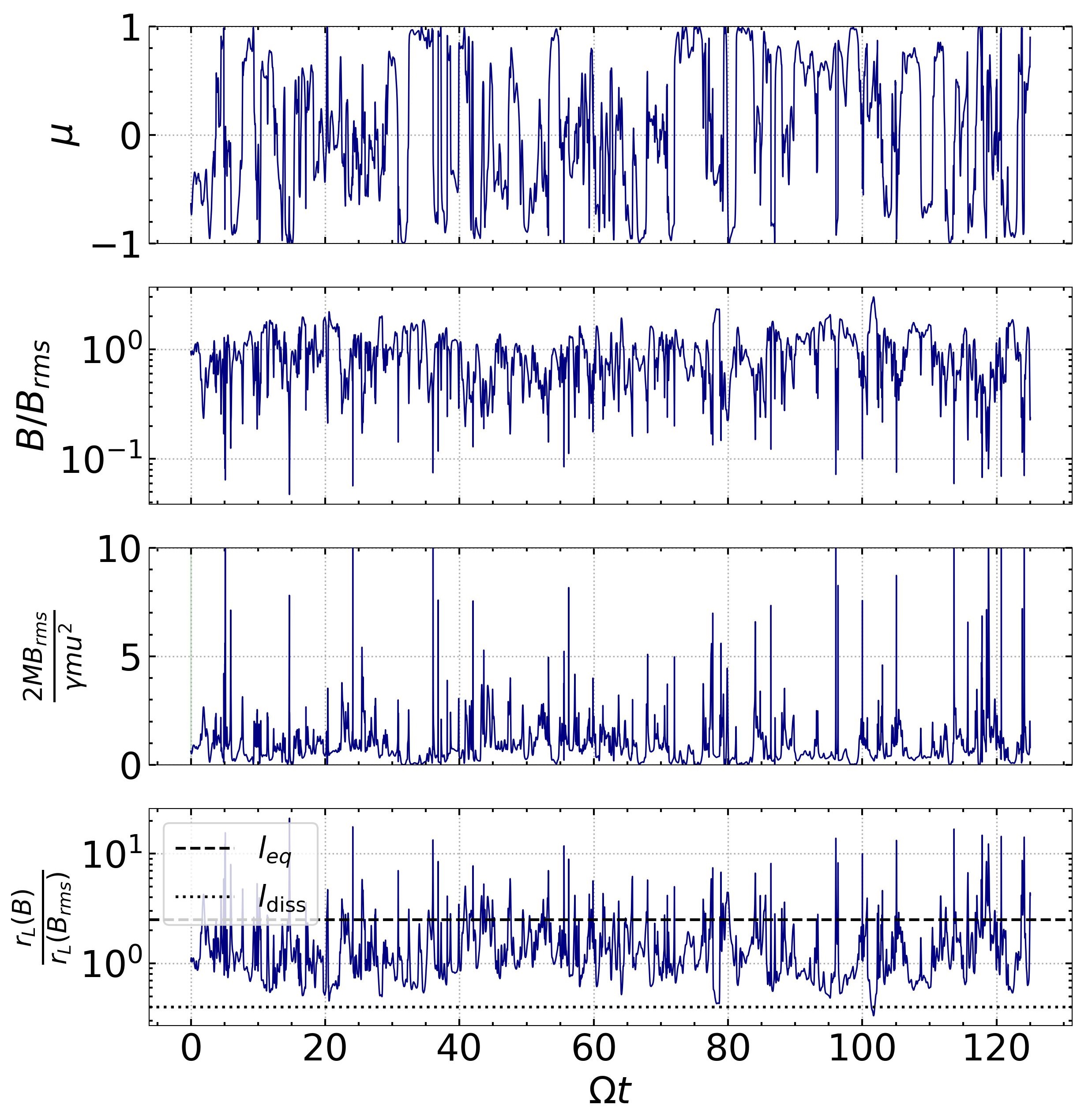}
  \caption{ Same as Fig. \ref{fig:CRM1_4} for a test particle with $r_L(B_\text{rms})=0.10L_\text{inj}$. The equipartition scale $l_\text{eq}$ and numerical dissipation scale $l_\text{diss}$ are indicated by horizontal dashed and dotted lines in the last panel. }
  \label{fig:CRM1_40}
\end{figure}

\begin{figure}[h!]
  \centering
    \hspace{-.0cm}\includegraphics[width=0.5\textwidth]{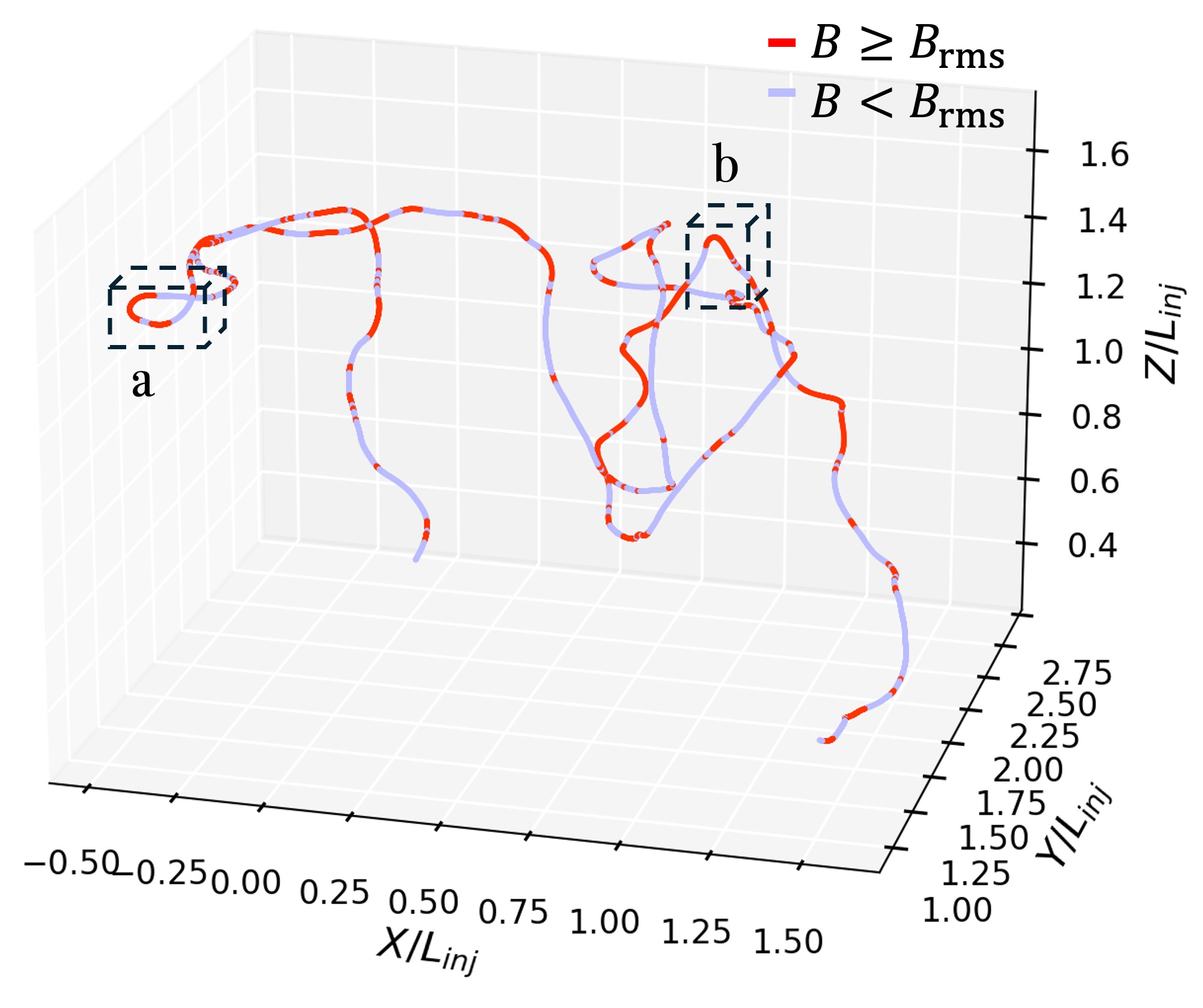}
  \caption{ Trajectory for the test particle shown in Fig. \ref{fig:CRM1_40}, color-coded by red when $B\geq B_\text{rms}$ and light blue when $B<B_\text{rms}$.  }
  \label{fig:CRM1_40_3d}
\end{figure}

\begin{figure*}
\gridline{\fig{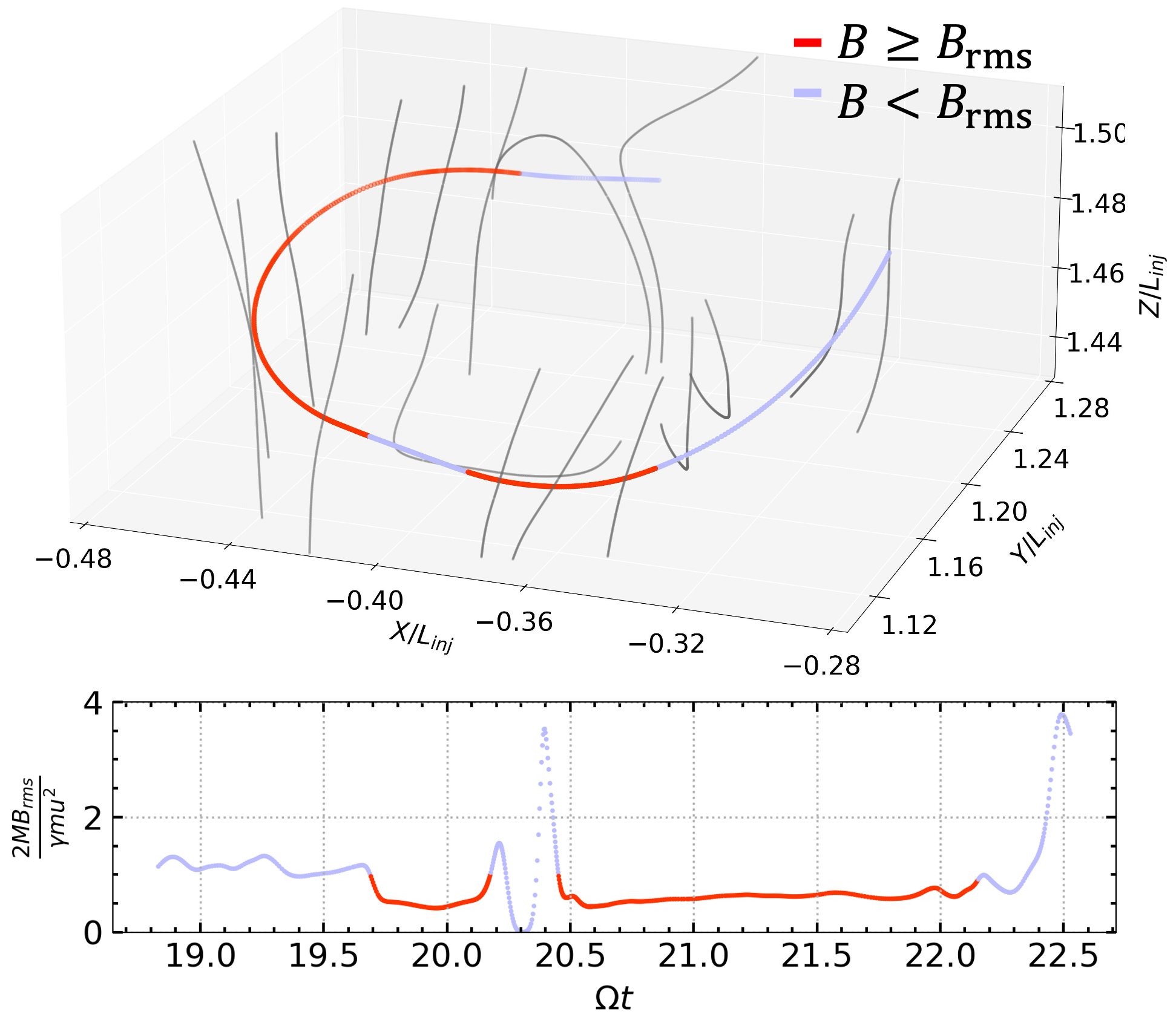}{0.48\textwidth}{(a)}
          \fig{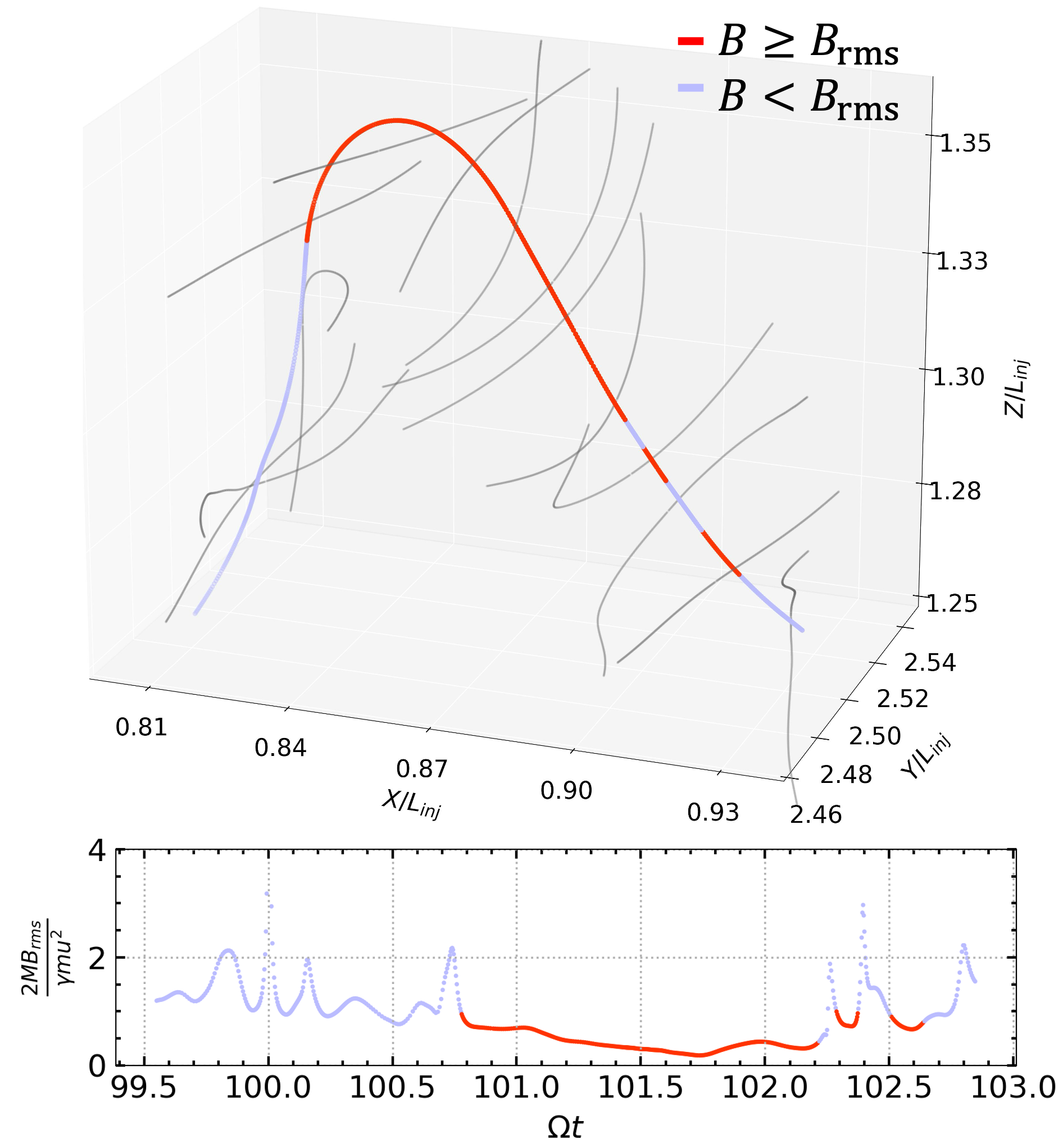}{0.45\textwidth}{(b)}
          }
\caption{Two zoom-ins on the ``a" (left) and ``b" (right) segments selected from the particle trajectory in Fig. \ref{fig:CRM1_40_3d}, with the same color code applied. The gray curves show the background magnetic field lines in the vicinity of the particle trajectory. Two lower panels show the normalized $M$ measured along the two segments of the trajectory. The particle performs a roughly half gyro-cycle in the strong-field region, causing a sharp turn of its trajectory, where $M$ is basically constant.}
\label{fig:CRM1_40_3d_two}
\end{figure*}

Next, we examine the diffusion of higher-energy CRs with $r_L(B_\text{rms})>l_\text{diss}$. As an example, Fig. \ref{fig:CRM1_40} presents the same measurements as in Fig. \ref{fig:CRM1_4} but for a higher-energy particle with $r_L(B_\text{rms})=0.10L_\text{inj}$. We find that different from the case with $r_L(B_\text{rms})<l_\text{diss}$, the diffusion of this particle is predominantly in the MMS regime and all four quantities exhibit significant variations through the entire time of study. 
Fig. \ref{fig:CRM1_40_3d} displays the corresponding particle  trajectory, which is color-coded by red when $B\geq B_\text{rms}$ and by light blue when $B< B_\text{rms}$. 
Similar to our finding in Fig. \ref{fig:CRM1_4_MMS}, we notice that the ``sharp turns" with significant change of the particle's moving direction (marked by ``a" and ``b" in Fig. \ref{fig:CRM1_40_3d} as examples) tend to happen when the local magnetic field is strong. To further look into the underlying mechanism, Fig. \ref{fig:CRM1_40_3d_two} shows the two zoom-ins on the ``a" and ``b" segments of the particle trajectory with surrounding magnetic field lines, 
as well as the time evolution of normalized $M$ corresponding to the two segments.
The stiff field lines in the strong field region and the bent field lines in the weak field region are consistent with the magnetic field structure shown in Fig. \ref{fig:BsliceM1} (a). 
We see that after entering the strong-field region, the particle performs a roughly half gyro cycle with the particle moving direction perpendicular to the magnetic field and a constant $M$. It then returns to the weak-field region where $M$ significantly varies. 
The limit case of mirroring when encountering the strong magnetic field can cause a sharp turn.
By contrast, the particle trajectory through the weak-field regions is relatively smooth (see Fig. \ref{fig:CRM1_40_3d}), indicative of the minor effect of weak magnetic fields on the diffusion of high-energy CRs.

As a brief summary, we find that the diffusion of low-energy CRs is inhomogeneous, with mirror and wandering diffusion preferentially taking place in strong fields and MMS diffusion in weak fields.
CRs are better confined in the mirroring and MMS regimes than in the wandering regime. High-energy CRs with their $r_L(B)$ too large to be confined within strong-field regions predominantly undergo the MMS diffusion. Notably, our results suggest that the incomplete gyrations of CRs, i.e., the limit case of mirroring, plays a more important role than the scattering in affecting the diffusion in the MMS regime (see Fig. \ref{fig:CRM1_4_MMS} and \ref{fig:CRM1_40_3d_two}).

\subsection{Time fractions of mirroring, wandering, and MMS at different CR energies}
\label{sec:statistical}
The comparison between Fig. \ref{fig:CRM1_4} and Fig. \ref{fig:CRM1_40} suggests that different mechanisms can dominate the diffusion of CRs at different energies. Based on their distinctive characteristics, we can separately identify the different diffusion regimes (Section \ref{sec:CRdiffusion}) and measure their corresponding time fractions. 
We consider $\Delta M/M<0.5$ for mirroring and wandering, and $\Delta M/M>0.5$ for MMS. 
Fig. \ref{fig:corM1} (a) presents the time fractions of mirroring and wandering measured at different CR energies. We find that they both drop to 0 as CR energy increases. Therefore, MMS becomes more important in affecting the CR diffusion toward higher energies. The minimum CR energy considered in this work is limited by the grid size. Based on the trend seen in Fig. \ref{fig:corM1} (a), we expect larger time fractions of mirroring and wandering toward lower CR energies.
The time fraction of wandering is always higher than that of mirroring for a given CR energy. Note that this result does not mean that wandering always tends to last longer than mirroring in each occurrence. It reflects the total time for wandering is longer than that of mirroring. 
As indicated in Fig. \ref{fig:CRM1_4}, mirror and wandering diffusion primarily happen in strong magnetic field regions. We find it also statistically true by measuring the $\langle B \rangle$, the averaged magnetic field strength sampled by the particles undergoing mirroring and wandering (see the gray circles in Fig. \ref{fig:corM1} (a)). The resulting $\langle B\rangle$ is always greater than $B_\text{rms}$ and increases with CR energy. 

Moreover, we can implement another independent measure of the relative importance of MMS. Unlike mirroring and wandering, as MMS causes the stochastic change of $\mu$, and thus loss of correlation of $\mu(t)$ with $\mu(t=0)$, we can measure the autocorrelation $R_{\mu\mu}(t)=\langle\mu(t=0)\mu(t)\rangle$ of $\mu$ over time averaged over all particles of the same energy, to estimate the relative importance of MMS. In Fig. \ref{fig:corM1} (b), the result shows that $R_{\mu\mu}$ decreases over time for all CR energies, due to the effect of the MMS, and it drops more rapidly for higher-energy CRs. This result suggests that the time fraction of MMS increases with CR energy, which is consistent with the finding in Fig. \ref{fig:corM1} (a). As CR energy increases, the autocorrelation is lost within one gyro-period. This implies that high-energy CRs do not undergo complete gyration in the MMS regime (see Fig. \ref{fig:CRM1_40_3d}). 

We should caution that the time fractions of different diffusion mechanisms measured here are for the particles that sample the entire volume with highly inhomogeneous distribution of magnetic fields (see Fig. \ref{fig:BsliceM1}). For instance, for CRs that preferentially sample the strong-field regions, the time fractions of mirroring and wandering are expected to be much higher than the results in Fig. \ref{fig:corM1} (a) (see Section \ref{sec:CRdiffusion}).

\subsection{Mean free path}
\label{sec:mfp}

\begin{figure*}[htb!]
\gridline{\fig{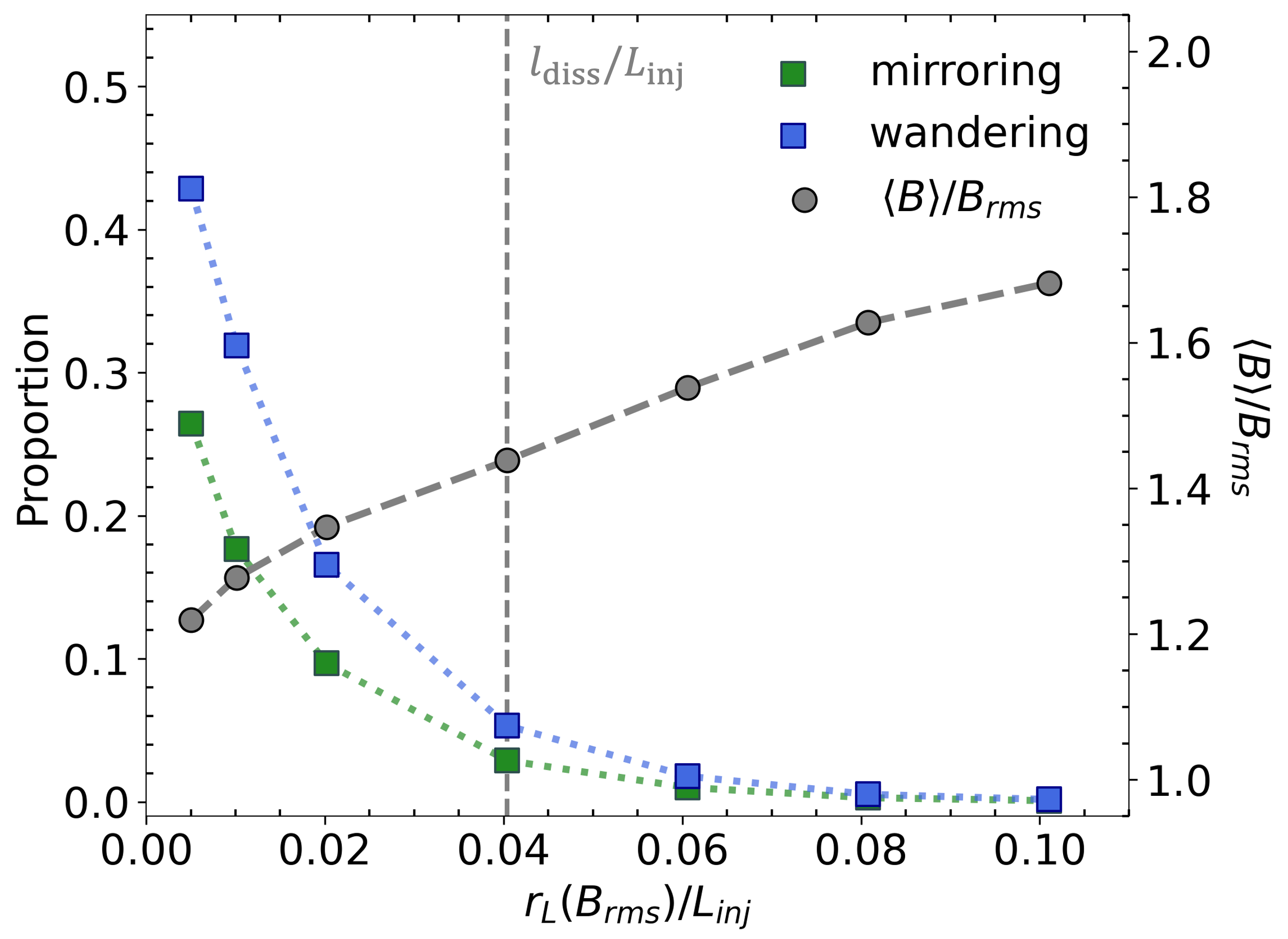}{0.47\textwidth}{(a)}
          \fig{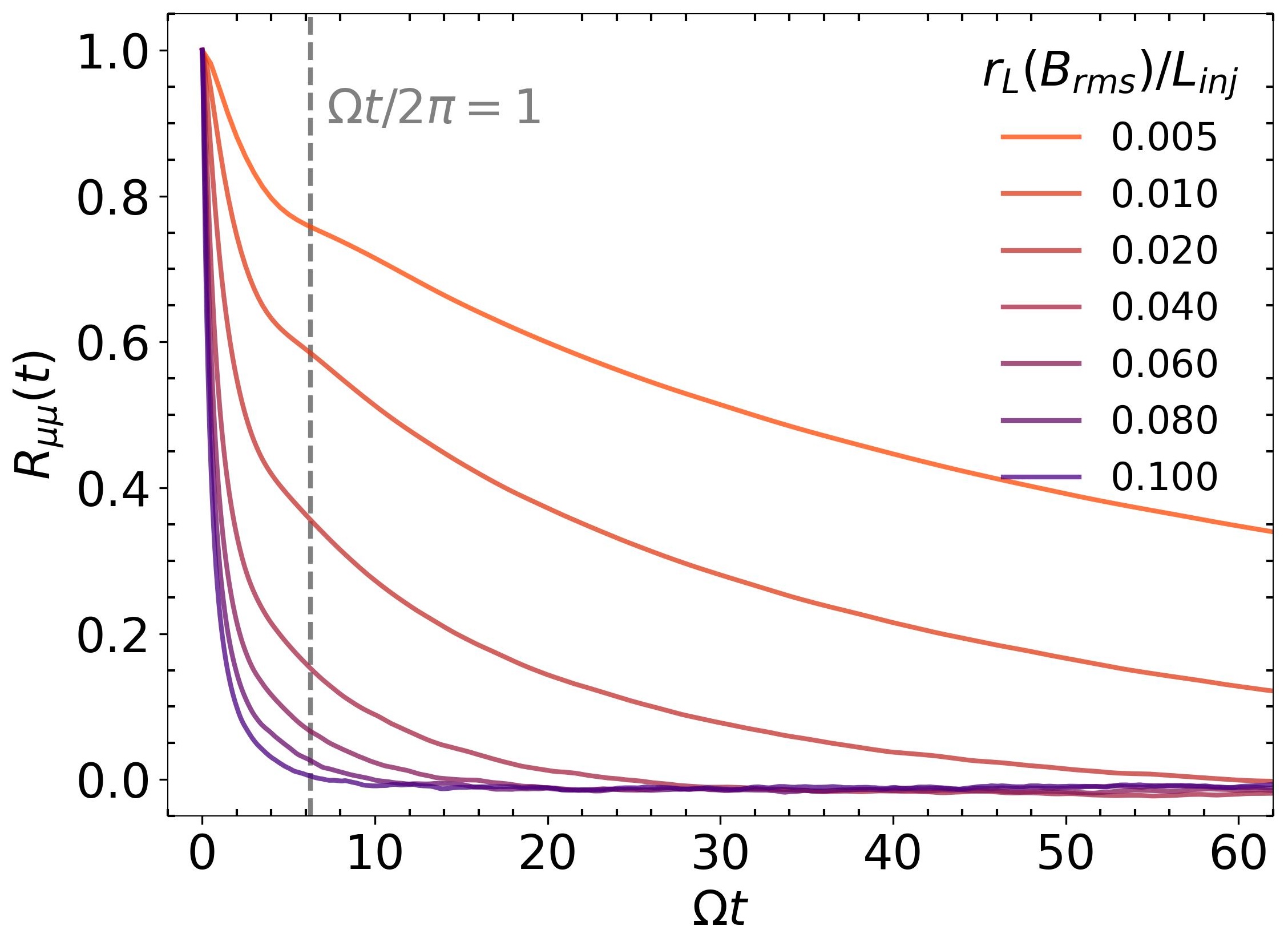}{0.47\textwidth}{(b)}
          }
\caption{(a) Time fractions of mirroring and wandering measured at different CR energies represented by green and blue squares, respectively. 
$\langle B\rangle/B_\text{rms}$ corresponding to the mirroring and wandering regimes are indicated by gray circles. (b) Autocorrelation of $\mu$ averaged over all particles as a function of $\Omega t$ for different CR energies.  }
\label{fig:corM1}
\end{figure*}

\begin{figure*}
\gridline{
    \fig{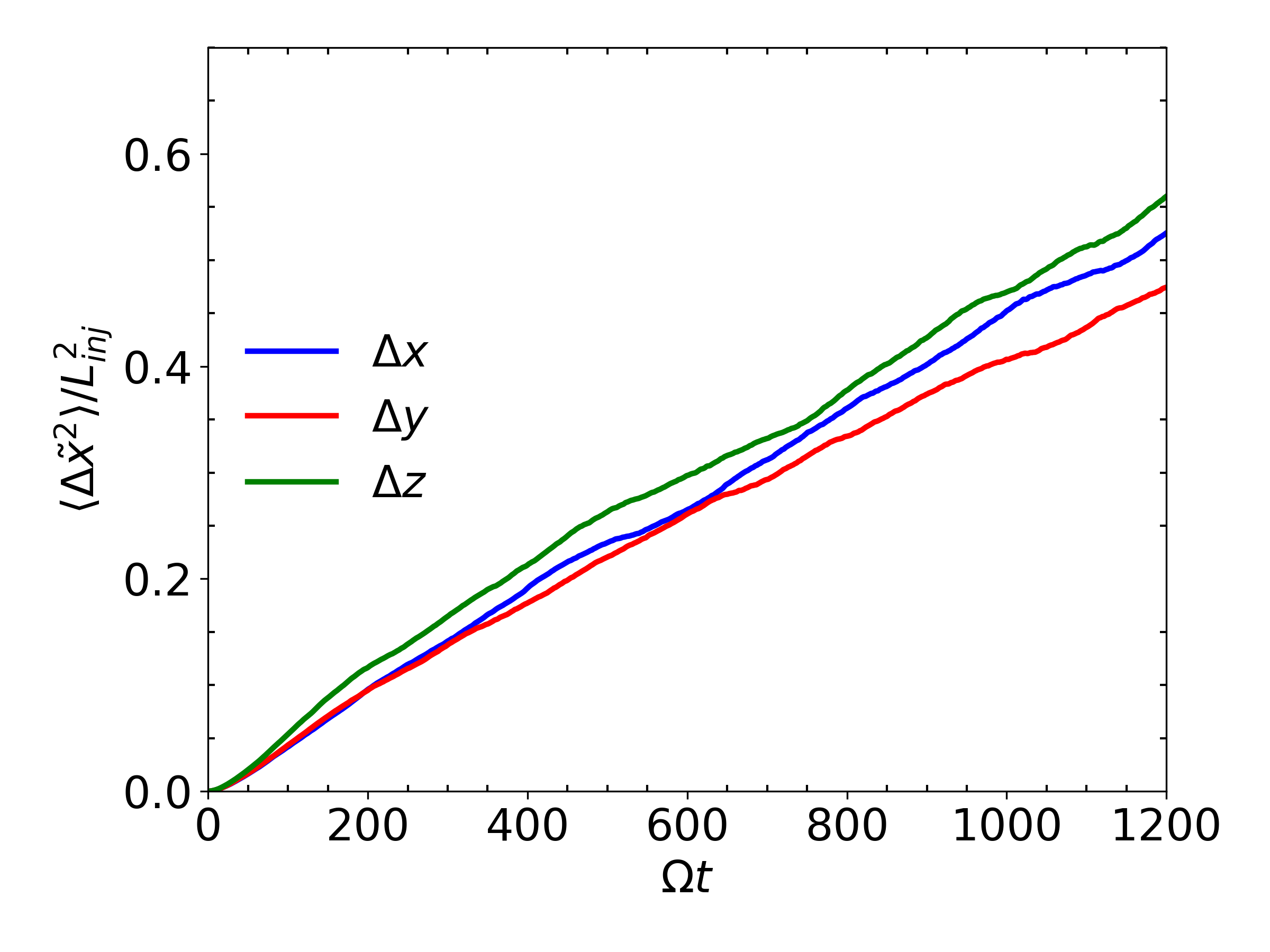}{0.45\textwidth}{(a) entire particle trajectories}
    \fig{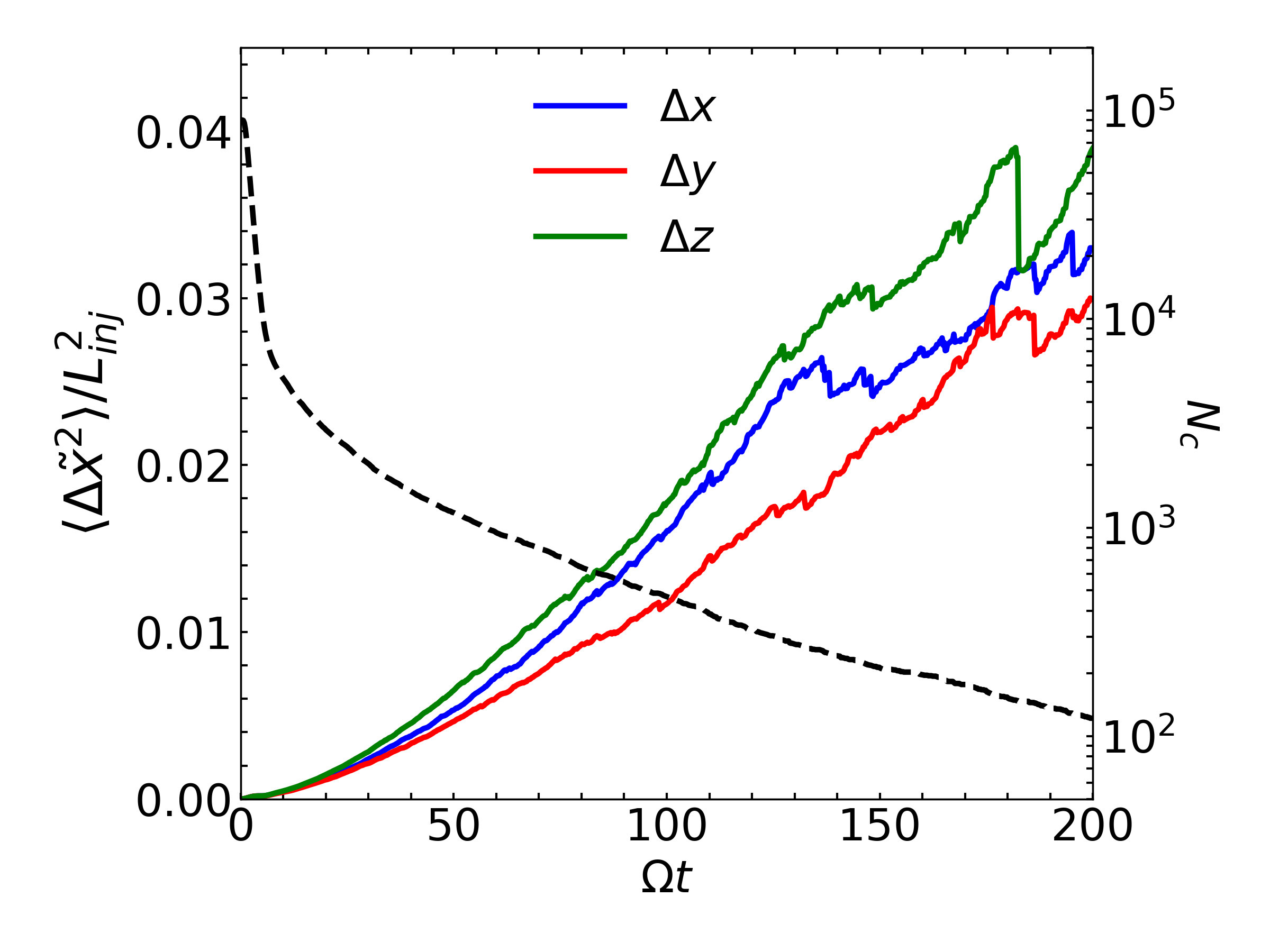}{0.45\textwidth}{(b) mirroring}
          }
\gridline{
          \fig{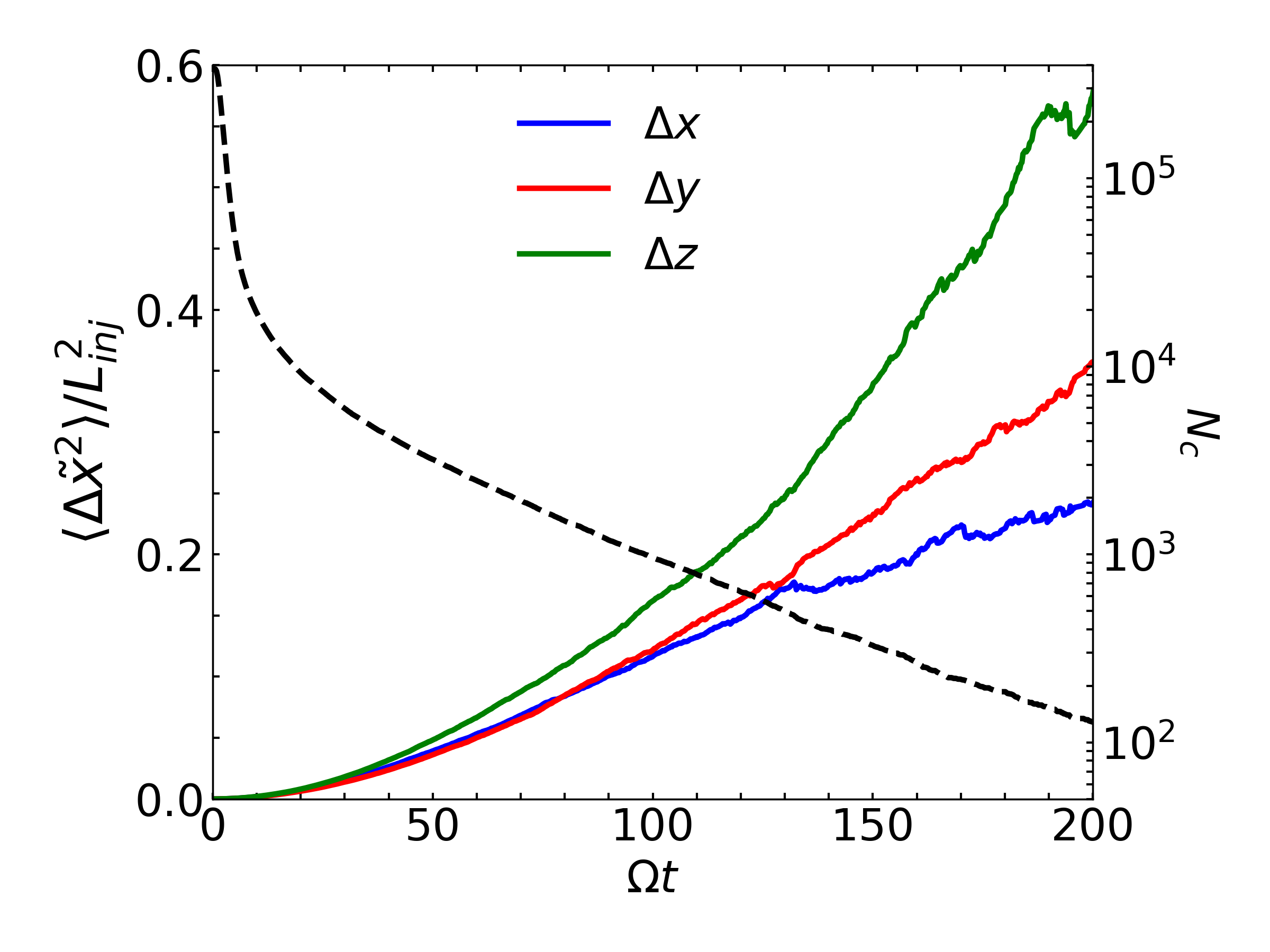}{0.45\textwidth}{(c) wandering}
          \fig{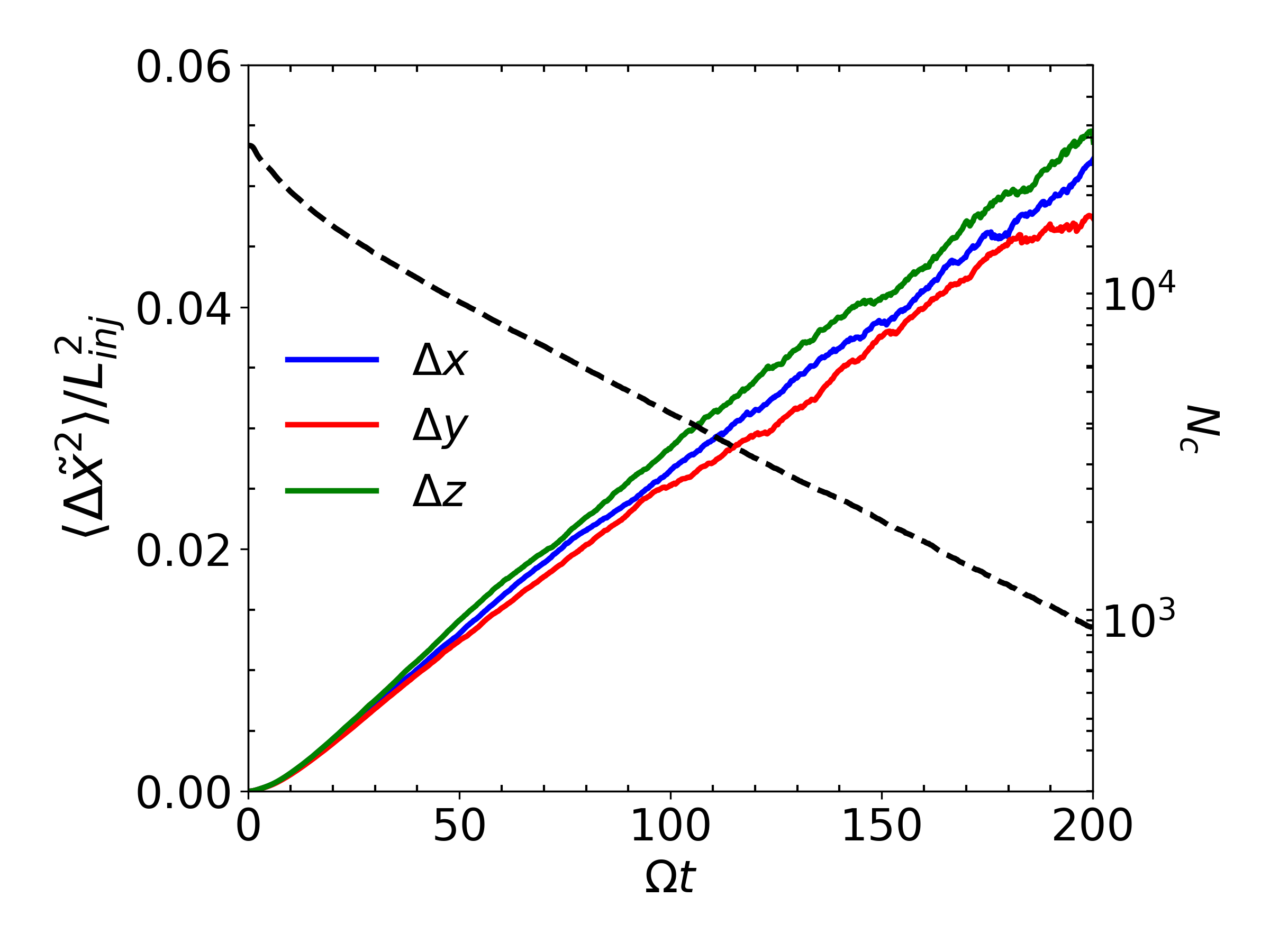}{0.45\textwidth}{(d) MMS}
          }
\caption{$\langle\Delta \Tilde{x}^2\rangle/L_\text{inj}^2$ vs. $\Omega t$ for entire particle trajectories in (a) and segments of trajectories corresponding to mirroring in (b), wandering in (c), and MMS in (d) at $r_L(B_\text{rms})=0.01L_\text{inj}$. Blue, red and green lines represent the mean squared displacement $\langle\Delta \Tilde{x}^2\rangle$ measured in x, y, and z directions. The black dash lines in (b), (c) and (d) represent the number of counts $N_c$ in each time bin. (a) has a constant sample size given by the total number of test particles. }
\label{fig:MFP_measure}
\end{figure*}

\begin{figure}[htb!]
  \centering
    \hspace{-.75cm}\includegraphics[width=0.5\textwidth]{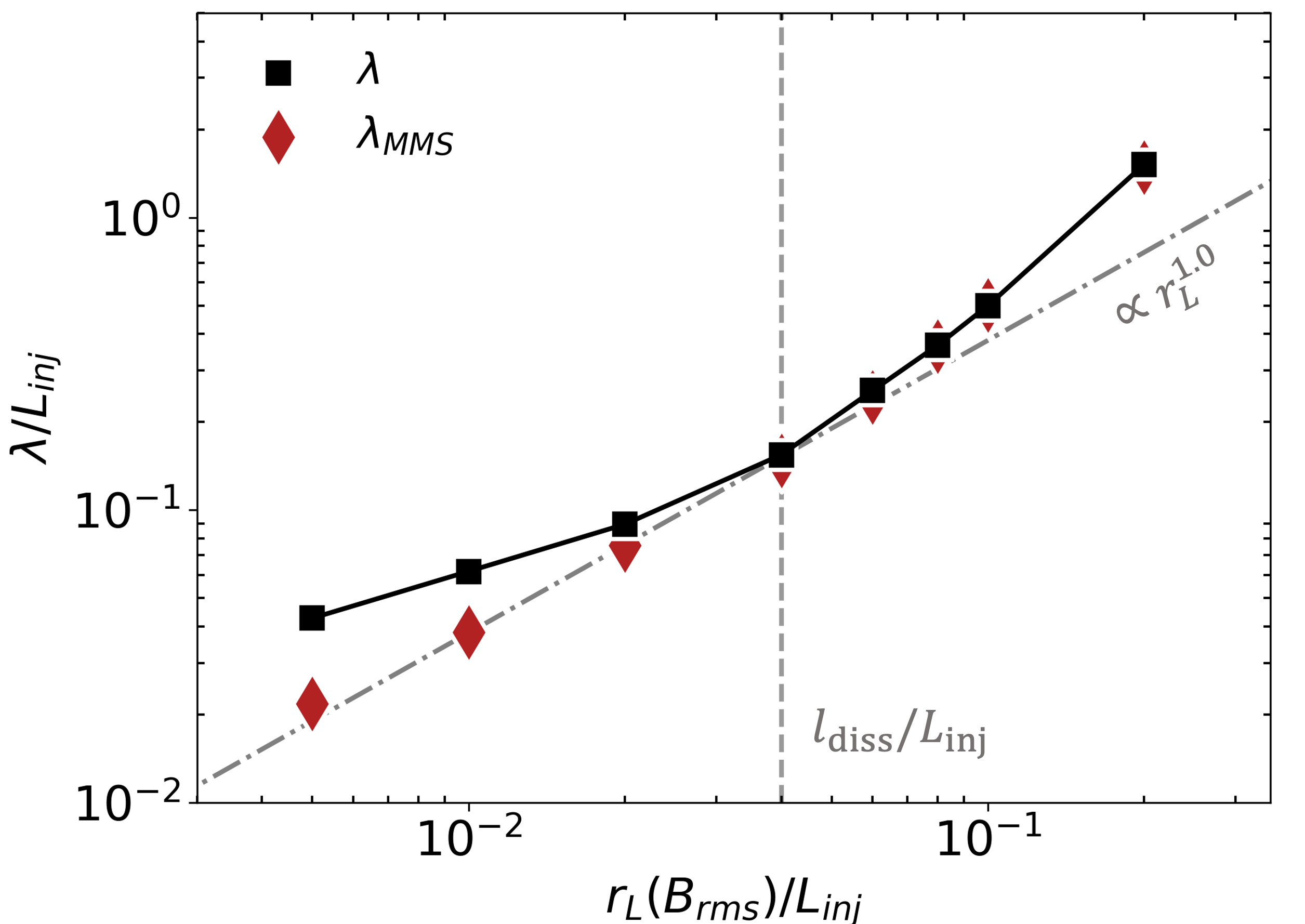}
  \caption{Overall $\lambda$ for entire CR trajectories (black squares) and $\lambda_\text{MMS}$ for CRs in the MMS regime (red diamonds) measured at different CR energies.
  Both $\lambda$ and $\lambda_\text{MMS}$ are averaged over x, y, and z directions. Vertical dashed line indicates the dissipation scale, and dash-dotted line indicates a linear dependence on energy as a reference.}
  \label{fig:mfpM1}
\end{figure}

To study the spatial confinement of CRs resulting from their diffusion, we perform the mean free path (MFP) measurement in the global reference frame for CRs with $r_L(B_\text{rms})=0.005L_\text{inj}$ to $0.2L_\text{inj}$. The energy range is chosen for the $r_L(B)$ of all particles to be within the box size but larger than the grid size.
The MFP is determined from calculating the spatial diffusion coefficient, 
\begin{equation}
    D = \frac{\langle \Delta \Tilde{x}^2\rangle}{2\Delta t}~,
\end{equation}
where $\Delta \Tilde{x}=\Delta x,\Delta y$, or $\Delta z$ is the displacement of a particle from its initial position measured in the corresponding direction. The average $\langle...\rangle$ is taken over all test particles at each energy and $\Delta t$ is the time corresponding to the displacement. 
When $D$ is a constant, that is, $\langle \Delta \Tilde{x}^2\rangle$ linearly depends on time, the MFP $\lambda$ is related to the diffusion coefficient by
\begin{equation}
    D = \frac{1}{3}u\lambda~. \label{eq:diffusion}
\end{equation}
With a constant $D$, CRs are in the diffusive regime. We then measure the slope of the $\langle \Delta \Tilde{x}^2\rangle$ vs. $\Delta t$ curve to estimate $D$ and $\lambda$. The total time of test particle simulation is sufficiently long such that the linear range is extended.

Fig. \ref{fig:corM1} clearly demonstrates the coexistence of mirroring, wandering and MMS especially for lower energy CRs. The role of each diffusion regime in confining CRs can be revealed by separately measuring its corresponding diffusion behavior. We measure $\langle \Delta \Tilde{x}^2\rangle/L_\text{inj}^2$ vs. $\Omega t$ for each diffusion regime. Here $\langle...\rangle$ is the average over all occurrences of a particular diffusion regime of all particles.
The $D$ and $\lambda$ measurements are only performed within the linear range of $\langle \Delta \Tilde{x}^2\rangle$ vs. $t$ and when there is sufficient number of counts of all occurrences for obtaining a statistically reliable result.

Fig. \ref{fig:MFP_measure} (a), (b), (c) and (d) show the results of $\langle \Delta \Tilde{x}^2\rangle/L_\text{inj}^2$ vs. $\Omega t$ for the entire particle trajectories, and for the segments of mirroring, wandering, and MMS, respectively, at low energy with $r_L(B_\text{rms})=0.01L_\text{inj}$ \footnote{In all cases, the displacement in z direction grows faster than that in x and y directions, probably because of the effect of the mean magnetic field in z direction. }. 
The number of counts $N_c$ for each diffusion regime within each time bin of size $1/\Omega$ is also included, which decreases over time due to the limited time span for a particle to remain in the same diffusion regime. The fluctuations in $\langle \Delta \Tilde{x}^2\rangle$ we see at $\Omega t\gtrsim 150$ in Fig. \ref{fig:MFP_measure} (b-d) are probably due to the insufficient sample size. 
Among the three diffusion regimes, we see that given the same time, the growth of $\langle \Delta \Tilde{x}^2\rangle$ for wandering CRs is one order of magnitude faster than that for mirroring and MMS, which have comparable growth of $\langle \Delta \Tilde{x}^2\rangle$ and stronger spatial confinement of CRs. This is expected, as the effective mean free path of wandering CRs is determined by $l_\text{eq}$ (\cite{brunetti2007compressible}).
We also find that only the MMS has an extended range with $\langle \Delta \Tilde{x}^2\rangle$ linearly dependent on time when CRs undergo normal diffusion. 
In both cases of mirroring and wandering, we see a superdiffusive behavior at early time. This can happen when the measured displacement is still within the MFP at the given CR energy. In Appendix \ref{sec:mirror_diffusion}, we measure $\langle \Delta \Tilde{x}^2\rangle$ for mirroring CRs at a lower energy with $r_L(B_\text{rms}) = 0.005L_\text{inj}$ and clearly see a diffusive behavior over a comparable displacement. It suggests that the mirror diffusion has an energy-dependent MFP (see LX21).
We note that as the displacements here are measured in the global reference frame, rather than along the local magnetic field, the observed superdiffusion at early time for mirroring and wandering may partly originate from the perpendicular superdiffusion of turbulent magnetic fields in the strong-field regions. However, we note that the perpendicular superdiffusion usually describes the growth of separation between a pair of CRs rather than their displacement
(\cite{xuyan2013}; \cite{lazarian2014superdiffusion}; \cite{yue2022superdiffusion}; ZX23). 

With contributions from all three diffusion regimes, 
$\langle \Delta \Tilde{x}^2\rangle$ measured for entire particle trajectories (Fig. \ref{fig:MFP_measure} (a)) grows linearly with time, indicative of an overall normal diffusion, while the superdiffusion at an early time due to the contributions from mirroring and wandering particles can be still seen.
This overall normal diffusion is the result of the stochastic transitions of CRs among three diffusion regimes. 

In Fig. \ref{fig:mfpM1}, we show the MFP $\lambda$ measured for the entire particle trajectories and the MFP $\lambda_\text{MMS}$ of CRs in the MMS regime at different energies. Under the consideration of the approximate isotropy of diffusion in both cases, we average $\lambda$ and $\lambda_\text{MMS}$ over three directions. 
We note that the differences in $\lambda$ (and $\lambda_\text{MMS}$) measured in three directions are smaller than the marker size, so they are neglected in Fig. \ref{fig:mfpM1}.
We see that $\lambda$ increases with CR energy, and the energy dependence becomes stronger toward higher energies with $r_L(B_\text{rms})>l_\text{diss}$.

At high energies with $r_L(B_\text{rms})>l_\text{diss}$, due to the dominance of MMS in diffusion, we see $\lambda =\lambda_\text{MMS}$. As demonstrated in Figs. \ref{fig:CRM1_40_3d} and \ref{fig:CRM1_40_3d_two}, the interaction with strong-field regions has a major effect on the diffusion of high-energy CRs. As the turning through a half-gyration can only happen when the local $r_L(B)$ is comparable to the size of the strong-field region, higher-energy CRs are preferentially affected  by stronger magnetic fields, with their $\lambda$ mainly determined by the spatial distribution of strong magnetic fields. Therefore, the strong energy dependence of $\lambda$ is likely a result of the smaller volume-filling factor of stronger magnetic fields (see Fig. \ref{fig:Bslice_cut}).

At lower energies with $r_L(B_\text{rms})<l_\text{diss}$, we see that $\lambda$ deviates from $\lambda_\text{MMS}$ due to the additional contributions from mirroring and wandering. The large and energy-independent effective MFP of wandering diffusion results in an overall larger $\lambda$ and its weaker energy dependence compared to $\lambda_\text{MMS}$. Given the increasingly importance of mirroring and wandering toward lower energies as indicated by Fig. \ref{fig:corM1} (a), we also expect a larger departure of $\lambda$ from $\lambda_\text{MMS}$ and the dominance of mirroring and wandering in determining $\lambda$ at further lower energies (with $r_L(B_\text{rms}) \lesssim 0.005L_\text{inj}$).
$\lambda_\text{MMS}$ at lower CR energies approximately increases linearly with energy. In Appendix \ref{sec:MMS}, we present an empirical model for explaining the energy dependence of $\lambda_\text{MMS}$.

We note that the overall $\lambda$ measured in Fig. \ref{fig:mfpM1} is averaged over the entire volume with contributions from all  three diffusion mechanisms. The MFP measured locally can be very different and depends on the dominant diffusion mechanism in the region sampled by the CRs (see Fig. \ref{fig:MFP_measure}).
We also caution that the negligible contributions from mirror and wandering diffusion at $r_L(B_\text{rms})> l_\text{diss}$ is likely to be a numerical artifact due to the limited numerical resolution (see Section \ref{sec:numeric}).

\section{Discussion}
\label{sec:discussion}

\subsection{Comparison with earlier studies}
\label{sec:compare}
CR diffusion in sub-Alfv\`enic MHD turbulence with a strong mean field has been extensively studied in the literature (e.g., \cite{chandran2000mirrorconfinement}; \cite{yan2002scatteringanisotropy}; \cite{beresnyak2011numerical}; \cite{xuyan2013}; \cite{cohet2016cosmic}; \cite{mertsch2020test}; \cite{yue2022superdiffusion}; ZX23). 
The gyroresonant scattering in sub-Alfv\'enic MHD turbulence faces the long-standing $90^\circ$ problem and cannot account for the observed diffusion of CRs by itself. Its combination with mirror diffusion naturally solves the $90 ^\circ$ problem and can explain both the suppressed diffusion in the vicinity of CR sources and faster diffusion in the diffuse interstellar medium (\cite{xu2021small}; ZX23; \cite{barreto2024cosmic}). 
Both mirror diffusion and scattering diffusion can happen everywhere in sub-Alfv\'enic MHD turbulence. The mirror diffusion takes place whenever $\mu$ is sufficiently small and $r_L(B)$ is smaller than $L_\text{inj}$. 

We find that in dynamo-amplified magnetic fields, mirror diffusion preferentially takes place in the strong-field regions when $r_L(B)$ is smaller than the sizes of these regions. The pitch-angle scattering is dominated by the interaction with small-scale weak and tangled magnetic fields. The wandering diffusion is uniquely seen in dynamo simulations as the correlation length of strong magnetic fields is smaller than the box size. Wandering diffusion can also affect streaming CRs (\cite{krumholz2020cosmic}).

Superdiffusive behavior is also found for streaming CRs (\cite{sampson2023turbulent}). However, we caution that the parallel and perpendicular superdiffusion have different physical origins. The latter originates from the perpendicular superdiffusion of turbulent magnetic fields (\cite{xuyan2013}; \cite{lazarian2014superdiffusion}; \cite{yue2022superdiffusion}; ZX23). The superdiffusion we see in the global reference frame (Fig. \ref{fig:MFP_measure} (b) and (c)) may have contributions from both.

More recent studies on CR diffusion in dynamo-amplified magnetic fields (e.g., \cite{lemoine2023particle}; \cite{kempski2023cosmic} focus on the mechanism of scattering by ``intermittent" small-scale magnetic field reversals.
Specifically, the term ``intermittency" was used to describe the power-law tail of the probability distribution function of magnetic field curvature (see also \cite{lubke2024towards}). 
In our work, we find that the diffusion of low-energy CRs is spatially inhomogeneous and identify three different diffusion regimes.
The scattering in the MMS regime is also seen in our simulations, but we find that the incomplete gyration in stronger fields with constant $M$ has a more significant effect on CR trajectories and diffusion compared to the curvature scattering that violates $M$ conservation.

\subsection{Numerical caveats}
\label{sec:numeric}
The slow diffusion of lower-energy CRs imposes constraints on the adoption of a snapshot of the dynamo simulation. Given the measured diffusion coefficient $D$, we can estimate the diffusion time $t_\text{D}=L^2/D$. Based on the smallest MFP $\lambda$ in Fig. \ref{fig:mfpM1}, we have the resulting ratio $t_\text{D}/t_\text{ed} = 3LV_L/\lambda c \approx 40 V_L/c \ll 1$ for nonrelativistic turbulence. 
Given $t_D \propto l^2$, and $t_\text{ed} \propto l^{2/3}$, where $l$ is the length scale, the diffusion time is always shorter than the eddy turnover time over all length scales. This justifies our approach using a turbulence snapshot.

In numerical simulations, due to the limited numerical resolution and the very limited range of $[l_\text{eq}, l_\text{diss}]$, the scattering efficiency and the time fraction of the MMS regime are significantly overestimated in strong-field regions. Given the limited range of scales, the scale-dependent anisotropy of MHD turbulence and its effect on suppressing scattering (\cite{chandran2000mirrorconfinement}; \cite{yan2002scatteringanisotropy}; \cite{xu2020trapping}) can hardly be numerically captured. By contrast, the mirroring rate is significantly underestimated as it increases with the parallel gradient of magnetic field, but the larger gradients toward smaller scales cannot be numerically resolved. In realistic situations with a sufficiently large range of scales, on scales larger than the physical $l_\text{diss}$, the relative importance between scattering and mirroring/wandering can be very different from our numerical results. 
\textbf{As fast modes can be subject to strong damping effects (e.g., \cite{brunetti2007compressible}), we compare the mirroring rate $\Gamma_m$ and the scattering rate $\Gamma_s$ corresponding to slow/pseudo-Alfv\'en modes. Their relative importance can be estimated as (\cite{xu2020trapping})}
\begin{equation}
    \frac{\Gamma_{m}}{\Gamma_{s}} \approx \frac{3}{8^{15/2}}\exp{(8)}\left(\frac{r_L(B_\text{rms})}{L_\text{inj}}\right)^{-\frac{3}{2}}(1-\mu^2)^{\frac{1}{2}}\mu^{-\frac{9}{2}}~,
\end{equation}
\textbf{which depends on the CR energy and pitch-angle. Under the consideration of $l_\text{diss}<r_L(B_\text{rms})\ll L_\text{inj}$, we expect that mirroring dominates over scattering. When $\mu$ approaches $1$, wandering is expected to be dominant.}
Therefore, the division by $l_\text{diss}$ shown in Fig. \ref{fig:mfpM1} is likely to be a numerical artifact due to the limited numerical resolution. 

The periodic boundary condition we impose may affect the measurement of the MFP when it approaches $L_\text{inj}$. Future work with higher resolution simulations will be carried out to further examine these numerical artifacts. 

\subsection{Astrophysical implications}
\label{sec:astro}
With the magnetic fields amplified by the small-scale turbulent dynamo in galaxy clusters (\cite{brunetti2011acceleration}; \cite{brunetti2014cosmic}; \cite{xu2020nonlinear}; \cite{kunz2022plasma}), this work has important implications on studying the diffusion and reacceleration of CR electrons in the ICM and explaining the extended radio halos (\cite{ensslin2011cosmic}; \cite{beduzzi2023exploring}; \cite{lazarian2023mirror}).
Given the correlation length of cluster magnetic fields of the order of 10 kpc (\cite{bonafede2010coma}), CR electrons with the radio observable energy range of $\sim$ 10 GeV (\cite{ensslin2011cosmic}) in $\mu$G-strength magnetic fields have $r_L/l_\text{eq} \sim 10^{-9}$, which is much lower than the minimum CR energy resolved in our simulations. Mirror and wandering diffusion is expected to be important for their diffusion.
Note that because of the inhomogeneity of CR diffusion at low energies, the dominant CR diffusion mechanism depends on the local magnetic field properties in the radio-emitting regions (\cite{hu2024synchrotron}).

\section{Conclusion}
\label{sec:conclusion}
By using test particle simulations, we study the CR diffusion in the magnetic field fluctuations amplified by the nonlinear turbulent dynamo, with a weak mean magnetic field. We summarize the main findings as the following. 

1. Dynamo-amplified magnetic fields have a highly inhomogeneous distribution, with smooth and coherent magnetic field lines in strong-field regions and tangled fields in weak-field regions. 
By contrast, sub-Alfv\'enic MHD turbulence with a strong mean magnetic field has a more homogeneous magnetic field distribution (e.g., \cite{cho2002compressible}). As CR diffusion strongly depends on the properties of turbulent magnetic fields, the inhomogeneity of dynamo-amplified magnetic fields fundamentally affects the CR diffusion and causes its difference compared to that in sub-Alfv\'enic MHD turbulence.

2. For low-energy CRs with $r_L(B_\text{rms})<l_\text{diss}$, we identified three different CR diffusion regimes, including the mirror, wandering, and MMS diffusion. The same mirror diffusion (LX21) has been earlier identified in sub-Alfv\'enic turbulence (ZX23; \cite{barreto2024cosmic}). The wandering diffusion predicted by \cite{brunetti2007compressible} is for the first time numerical demonstrated. The mirror and wandering  diffusion plays a more important role in affecting the overall CR diffusion behavior, with increasing time fractions toward lower CR energies. Both diffusion mechanisms preferentially take place in the strong-field regions, while the MMS diffusion is seen in weak-field regions. Mirror and MMS diffusion is much slower than the wandering diffusion. 

3. Higher Energy CRs with $r_L(B_\text{rms})>l_\text{diss}$ predominantly undergo the MMS diffusion, irrespective of the magnetic field inhomogeneity. However, we caution that our measurement on the relative importance of different diffusion regimes at $r_L(B_\text{rms})>l_\text{diss}$ can be severely affected by the limited numerical resolution (see Section \ref{sec:numeric}). Compared with earlier studies on the curvature scattering by small-scale magnetic field reversals in dynamo-amplified magnetic fields (e.g., \cite{lemoine2023particle}; \cite{kempski2023cosmic}), we also identify the similar scattering in the MMS regime, but our results suggest that the incomplete particle gyration, i.e., the limit case of mirroring, has a dominant effect on the MMS diffusion.

4. The overall MFP $\lambda$ increases with CR energy and shows a stronger energy dependence at higher energies. 
At lower energies, $\lambda$ has a weaker energy dependence compared to $\lambda_\text{MMS}$ because of the contribution from the energy-independent wandering diffusion. At higher energies, $\lambda$ becomes equal to $\lambda_\text{MMS}$.

Our findings can have important implications on studying the CR diffusion and reacceleration in the ICM. It is worth noting that the overall MFP measurement in this work is based on a result averaged over the entire volume of turbulent magnetic fields. Given the inhomogeneous dynamo-amplified magnetic fields in galaxy clusters (\cite{hu2024synchrotron}) and the inhomogeneous diffusion of CRs found in this work, an application of this study to observations requires understanding of the local magnetic field structures of the radio-emitting regions in galaxy clusters. 

\acknowledgments

We are grateful to the anonymous referee for thoughtful comments. We thank Martin Lemoine and Philipp Kempski for helpful discussions. We acknowledge the support for this work from SDSC Expanse CPU in SDSC through the allocation PHY230032 from the Advanced Cy-berinfrastructure Coordination Ecosystem: Services \& Support (ACCESS) program, which is supported by National Science Foundation grants No. 2138259, 2138286, 2138307, 2137603, and 2138296. We acknowledge University of Florida Research Computing for providing computational resources and support that have contributed to the research results reported in this publication.
This work was performed in part at the Aspen Center for Physics, which is supported by National Science Foundation grant PHY-2210452 and Durand Fund. This research was supported in part by grant NSF PHY-2309135 to the Kavli Institute for Theoretical Physics (KITP). SX also acknowledges the inspiring discussion with the participants of the program on ``Turbulence in Astrophysical Environments" at KITP. We acknowledge the support from NASA ATP award 80NSSC24K0896.

\bibliographystyle{aasjournal}
\bibliography{bibli}

\appendix
\section{Distributions of weak and strong magnetic fields in nonlinear turbulent dynamo}
\label{sec:bslice}
\begin{figure*}[htb!]
\gridline{\fig{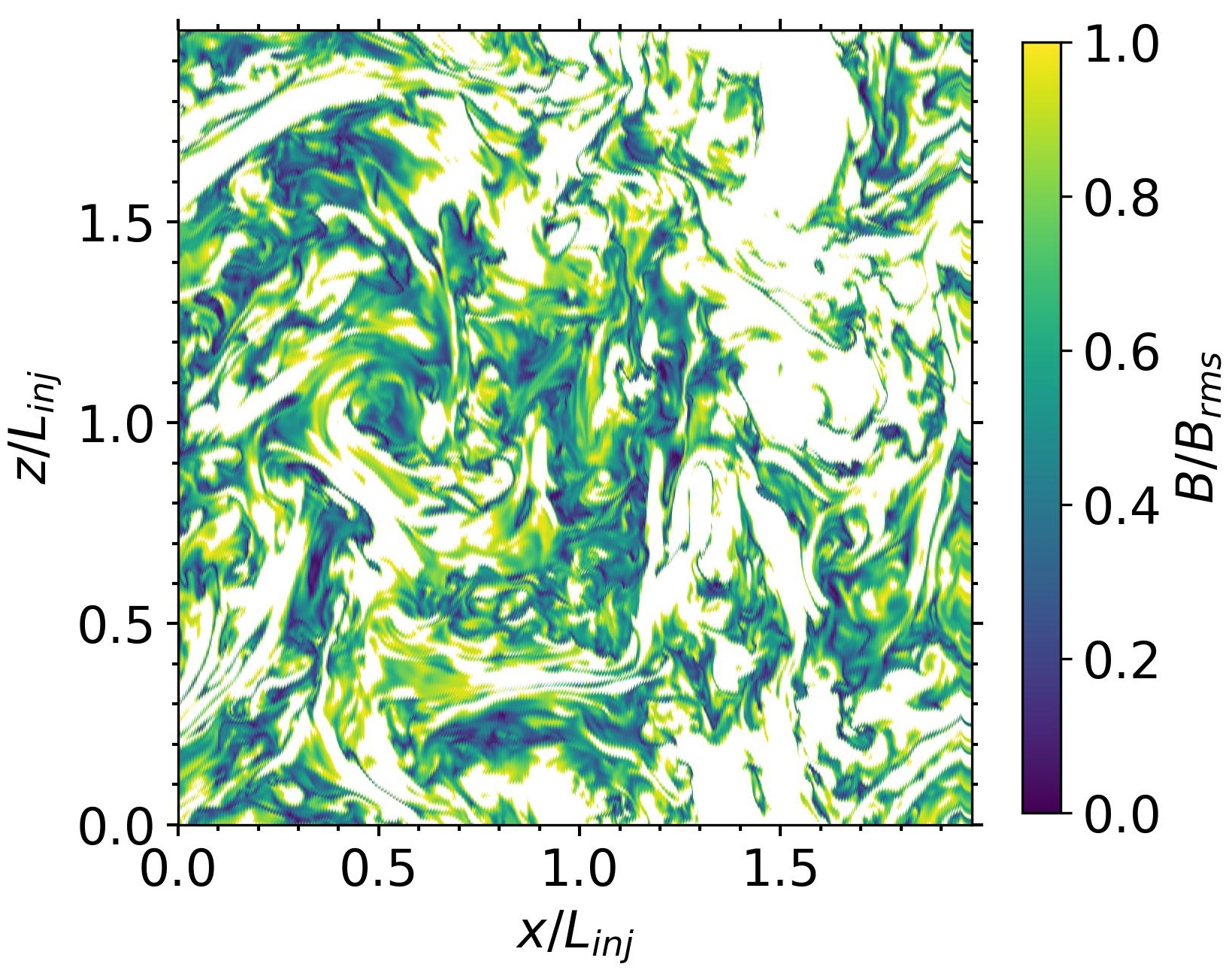}{0.45\textwidth}{(a) ``Weak" magnetic fields}
          \fig{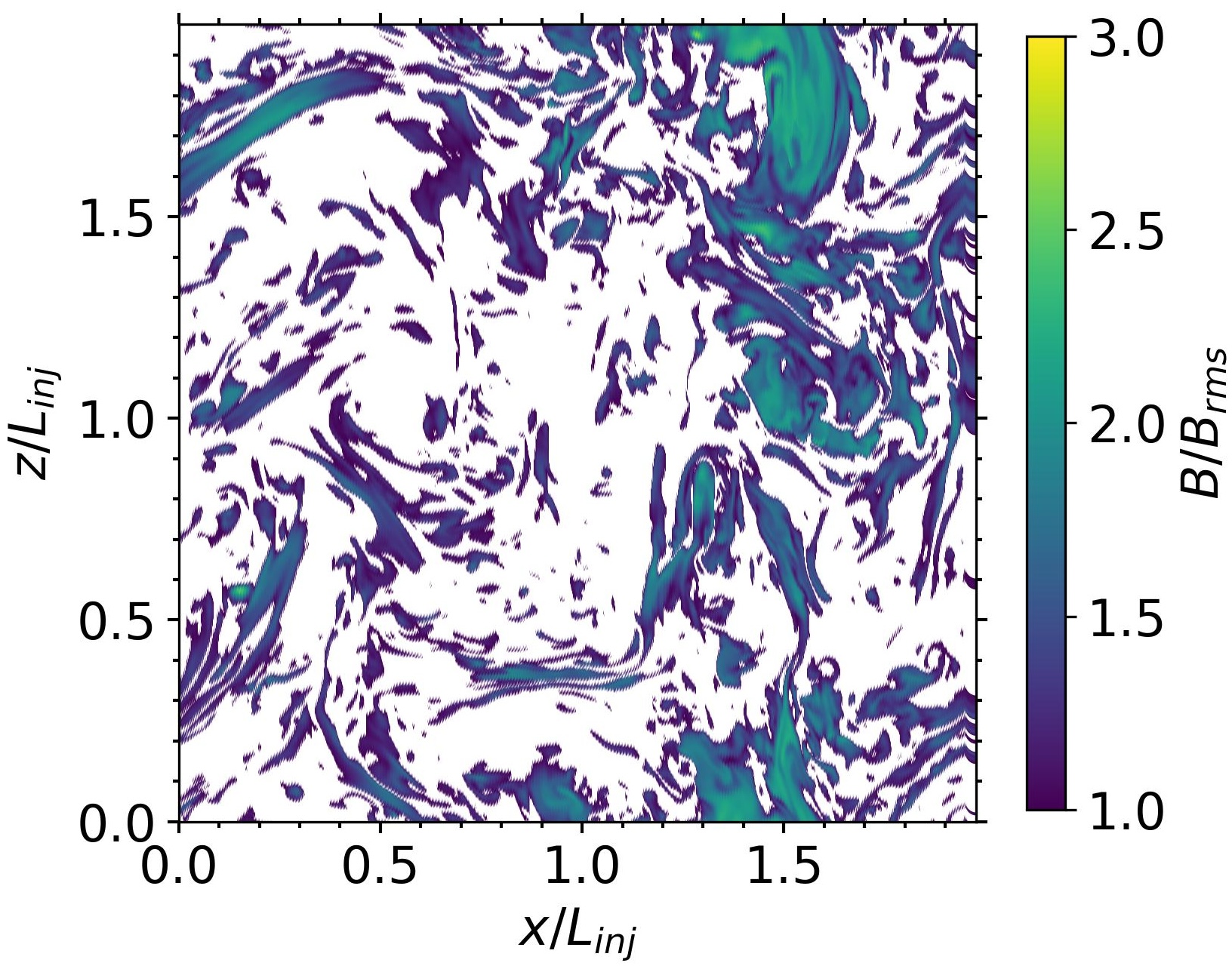}{0.45\textwidth}{(b) ``Strong" magnetic fields}
          }
\caption{ Same as Fig. \ref{fig:BsliceM1} (a) but for $B<B_\text{rms}$ in (a) and $B\geq B_\text{rms}$ in (b).  }
\label{fig:Bslice_cut}
\end{figure*}

In Fig. \ref{fig:Bslice_cut}, we separately present the distributions of the ``weak" fields with $B<B_\text{rms}$ and  the ``strong" fields with  $B\geq B_\text{rms}$. The spatial inhomogeneity is still clearly seen, especially for the ``weak" fields due to its large strength range (see Fig. \ref{fig:BdisM1}). The ``weak" fields are relatively more space-filling, while the ``strong" fields are characterized by both large-scale patches of strong coherent magnetic fields and small-scale structures at $B \sim B_\text{rms}$.

\section{Mirror diffusion at a lower CR energy}
\label{sec:mirror_diffusion}
\begin{figure}[H]
  \centering
    \hspace*{-0cm}\includegraphics[width=0.5\textwidth]{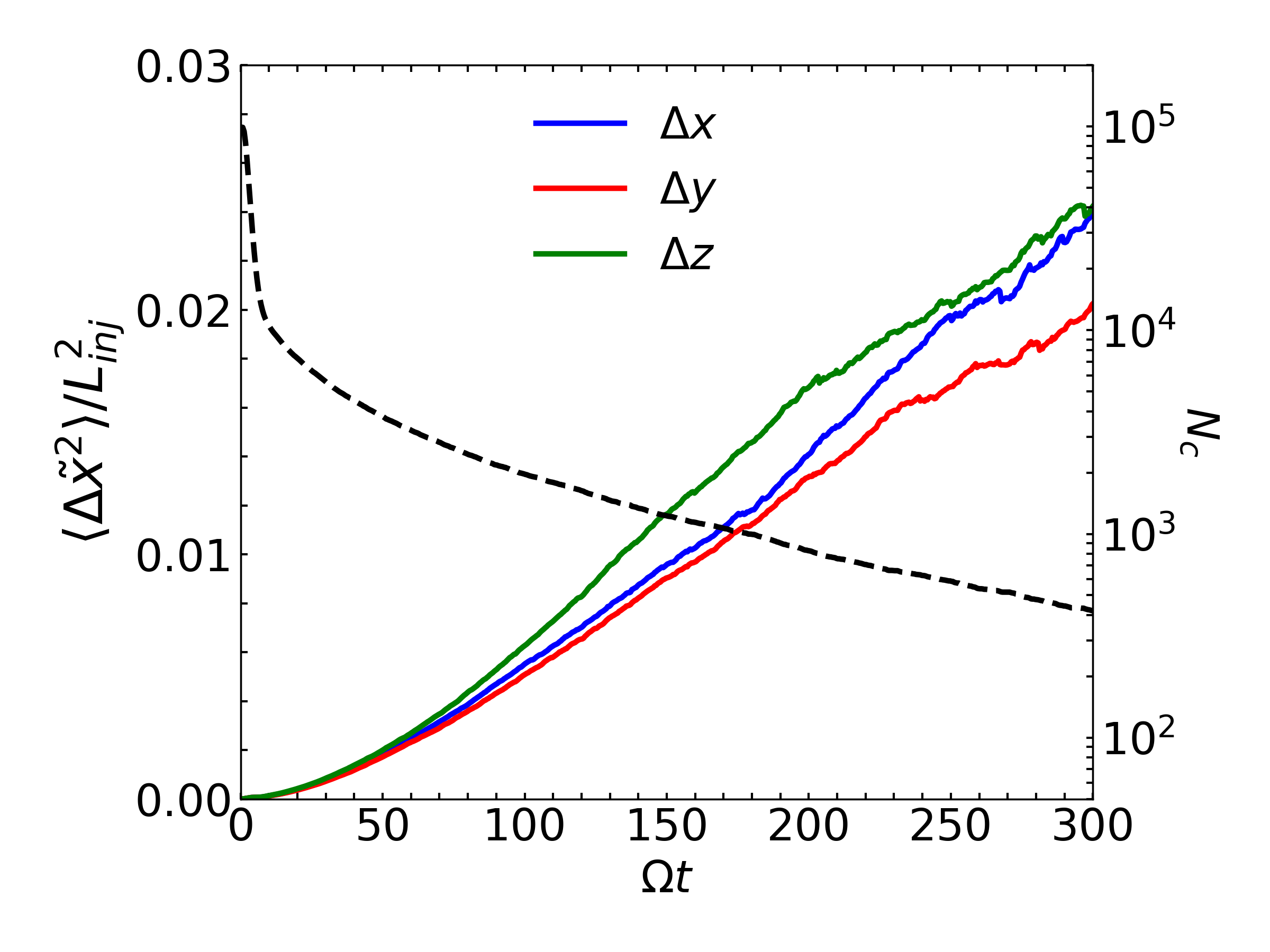}
  \caption{Same as Fig. \ref{fig:MFP_measure} (b) but for CRs with $r_L(B_\text{rms}) = 0.005L_\text{inj}$. }
    \label{fig:mirror}
\end{figure}

Here as shown in Fig. \ref{fig:mirror}, we measure $\langle\Delta \Tilde{x}^2\rangle/L_\text{inj}^2$ vs. $\Omega t$ for mirroring CRs with $r_L(B_\text{rms}) = 0.005L_\text{inj}$. An asymptotic range with normal diffusion can be seen from $\langle\Delta \Tilde{x}^2\rangle>0.005L_\text{inj}^2$. The calculated MFP is $\lambda\approx 0.03L_\text{inj}$ in all directions.

\section{Modeling of MFP in the MMS regime}
\label{sec:MMS}

\begin{figure}[htb!]
  \centering
    \hspace*{-0cm}\includegraphics[width=0.5\textwidth]{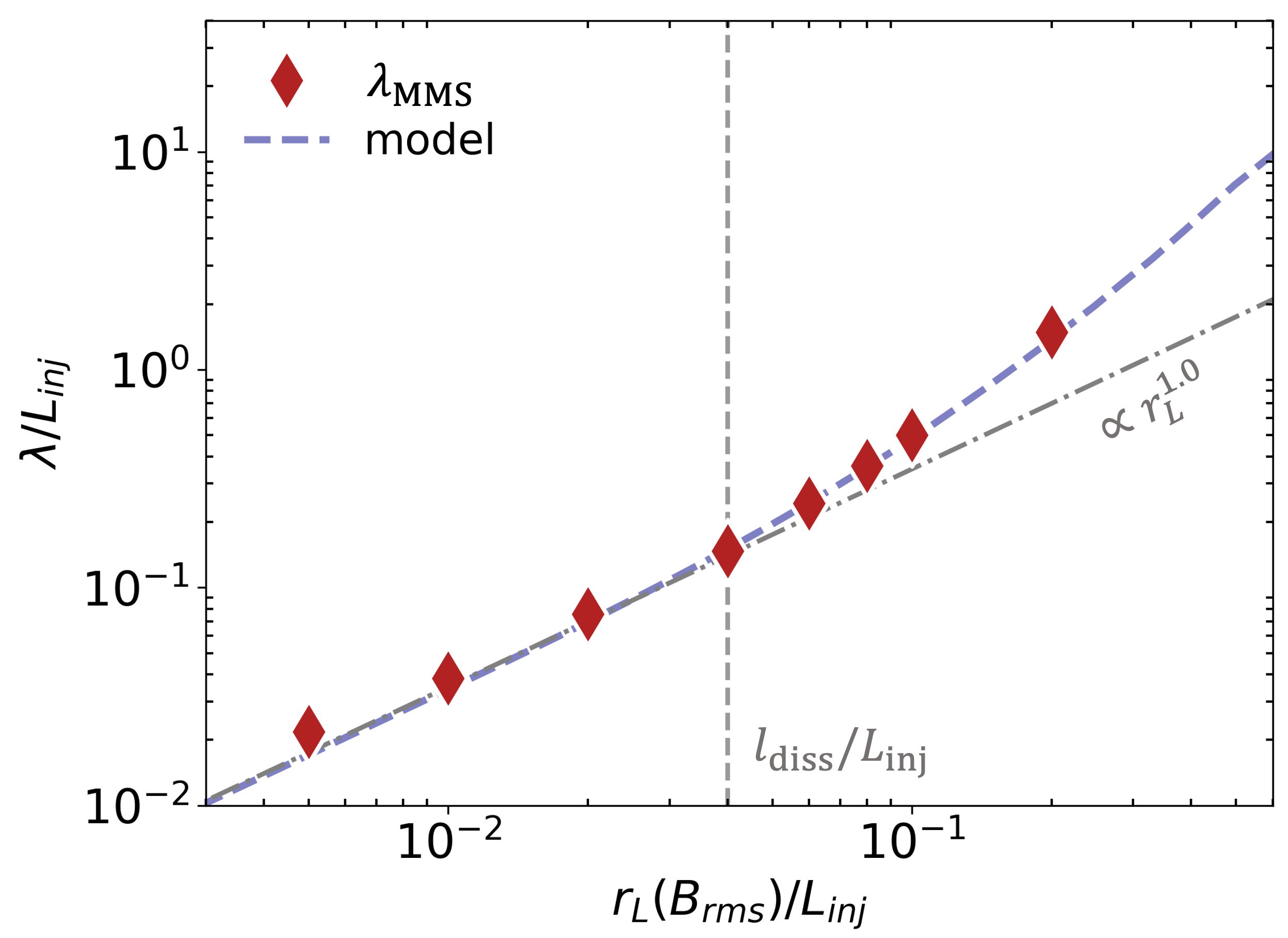}
  \caption{Comparison between the measured (red diamonds) and modeled (blue dashed line) $\lambda_\text{MMS}$. The latter is calculated using Eq. \ref{eq:MMS} with $\mathcal{C} = 3.4$ }
    \label{fig:MMS}
\end{figure}

The interaction with stronger fields at scales comparable to the local $r_L(B)$ dominates the diffusion of higher-energy CRs.
This finding motivates a definition of modified magnetic coherence length $\ell_c$, assuming magnetic fields at the scale smaller than $r_L$ are smoothed out,
\begin{equation}
    \ell_c(r_L)=\frac{\int_1^{1/r_L}\frac{1}{k}E_B(k)dk}{\int_1^{1/r_L}E_B(k)dk}~.
    \label{eq:coherence}
\end{equation}
This length approaches to the conventional magnetic coherence length $l_c$ (\cite{cho2009characteristic}) as $r_L\rightarrow l_\text{diss}$. Based on $E_B(k)$ measured in Fig. \ref{fig:spectrum} at $t/t_\text{ed}=2.2$, we obtain $\ell_c(r_L)$. We then model $\lambda_\text{MMS}$ with the following expression, 
\begin{equation}
    \lambda_\text{MMS}(r_L) = \mathcal{C} r_L \frac{\ell_c(r_L)}{l_c}~,\label{eq:MMS}
\end{equation}
where $r_L$ can be estimated by $r_L(B_\text{rms})$ and $\mathcal{C}$ is a fitting parameter. By comparing the above expression with our numerical measurements, we find $\mathcal{C} \approx 3.4$. The measured and modeled $\lambda_\text{MMS}$ are shown in Fig. \ref{fig:MMS}.
The stronger energy dependence of $\lambda_\text{MMS}$ at higher energies is due to the smaller volume filling fraction of stronger fields (see Fig. \ref{fig:Bslice_cut} (b)).

\end{CJK*}
\end{document}